\documentclass[useAMS,usenatbib]{mnras}
\usepackage{graphicx,lscape,amssymb}
\usepackage{natbib}
\usepackage{fancyhdr}
\usepackage{subfigure}
\usepackage{longtable}
\usepackage{rotating}
\usepackage{calc}
\usepackage{fancyhdr}
\usepackage{float}
\usepackage{enumerate}
\usepackage{sidecap}

\usepackage{amsmath}
\usepackage{subfigure}
\usepackage{booktabs, dcolumn}
\usepackage[dvipsnames]{xcolor}
\usepackage{hyperref}
\newcommand{\finsize}{$486\,641$}
\newcommand{\Teff}{\mbox{$T_{\mathrm{eff}}$}} 

\newcommand{\Pwd}{\mbox{$P_{\mathrm{WD}}$}}
\usepackage{soul}

\newcommand{\bp}{\mbox{$G_{\rm BP}$}}
\newcommand{\rp}{\mbox{$G_{\rm RP}$}}
\newcommand{\Gabs}{\mbox{$G_{\rm abs}$}}

\title[A \textit{Gaia} DR 2 catalogue of white dwarfs]{A {\em Gaia} Data Release 2 catalogue of white dwarfs and a comparison with SDSS}

\author[Gentile Fusillo et al.]  {Nicola Pietro Gentile Fusillo,$^{1}$\thanks{E-mail:
    N.Gentile-Fusillo@warwick.ac.uk} Pier-Emmanuel Tremblay,$^{1}$ Boris T. G\"ansicke,$^1$
  \newauthor{Christopher J. Manser,$^1$ Tim Cunningham,$^1$ Elena Cukanovaite,$^1$ Mark Hollands,$^1$}
  \newauthor{Thomas Marsh,$^1$ Roberto Raddi,$^2$ Stefan Jordan,$^3$ Silvia Toonen,$^4$}
  \newauthor{Stephan Geier,$^5$ Martin Barstow,$^6$ and Jeffrey D. Cummings$^7$}
  \\
  $^{1}$Department of Physics, University of Warwick, CV4 7AL, Coventry, UK\\
  $^{2}$Dr. Karl Remeis-Observatory, Friedrich-Alexander University Erlangen-Nuremberg, Sternwart-str. 7, 96049 Bamberg, Germany\\
  $^{3}$Astronomisches Rechen-Institut, Zentrum f\"ur Astronomie der Universit\"at Heidelberg, D-69120 Heidelberg, Germany\\
  $^{4}$Anton Pannekoek Institute for Astronomy, University of Amsterdam, 1090 GE Amsterdam, The Netherlands\\
    $^{5}$University of Potsdam, Institute of Physics and Astronomy, Karl-Liebknecht-Strasse 24/25, 14476 Potsdam, Germany\\
  $^{6}$University of Leicester, Leicester Institute of Space \& Earth Obs., Physics Building, University Road, Leicester LE1 7RH, UK\\
    $^{7}$Center for Astrophysical Sciences, Johns Hopkins University, Baltimore, MD 21218, USA}
  
\begin{document}
\maketitle

\label{firstpage}
\begin{abstract}
We present a catalogue of white dwarf candidates selected from the second data release of \textit{Gaia} (DR2).
We used a sample of spectroscopically confirmed white dwarfs from the Sloan Digital Sky Survey (SDSS) to map the entire space spanned by these objects in the \textit{Gaia} Hertzsprung-Russell diagram. We then defined a set of cuts in absolute magnitude, colour, and a number of \textit{Gaia} quality flags to remove the majority of contaminating objects. Finally, we adopt a method analogous to the one presented in our earlier SDSS photometric catalogues to calculate a probability of being a white dwarf (\Pwd) for all \textit{Gaia} sources which passed the initial selection.
The final catalogue is composed of \finsize\ stars with calculated \Pwd\ from which it is possible to select a sample of $\simeq 260\,000$ high-confidence white dwarf candidates in the magnitude range $8<G<21$. By comparing this catalogue with a sample of SDSS white dwarf candidates we estimate an upper limit in completeness of 85 per cent for white dwarfs with $G \leq 20$\,mag and  \Teff\ $> 7000$\,K, at high Galactic latitudes ($|b|>20^{\circ}$). However, the completeness drops at low Galactic latitudes, and the magnitude limit of the catalogue varies significantly across the sky as a function of \textit{Gaia}'s scanning law. We also provide the list of objects within our sample with available SDSS spectroscopy. We use this spectroscopic sample to characterise the observed structure of the white dwarf distribution in the H-R diagram.
\end{abstract}
\section{Introduction}
All stars with main sequence masses $\lesssim8-10$ $\mathrm{M}_\odot$ \citep{ibenetal97-1,dobbieetal06-1} share the same common fate: they will one day evolve into white dwarfs, dense stellar embers destined to cool over billions of years \citep{fontaine01,althaus10}. This broad mass range includes over 90 per cent of all stars in the Galaxy. This makes white dwarfs significant contributors to the global stellar population and, thanks to their well defined cooling rates, accurate tracers of the formation and evolution of the Milky Way (e.g., \citealt{winget87,torresetal05-1, tremblayetal14-1}). The diagnostic potential of the Galactic white dwarf population can only be fully exploited once we have large, homogeneous, and well-defined samples of white dwarfs. Given the intrinsic low luminosities and relatively high proper motions of stellar remnants, these samples have been historically challenging to assemble.  

The fundamental properties of white dwarfs (mass, cooling age, atmospheric and internal composition) can be determined from spectroscopic, photometric or asteroseismic analyses \citep{bergeronetal92-1,bergeron01,koester09,bergeronetal11-1,tremblayetal13-1,romero17,gia18}. These parameters are essential to constrain and calibrate stellar evolution theory. Important examples are the mass loss on the AGB (intimately linked to the initial-to-final mass relation, e.g.,  \citealt{Weidemann77,dobbie06,williams09,kalirai14,romero15,cummings16}), internal rotation profiles and loss of angular momentum \citep{charpinet09,hermes17}, and fundamental nuclear reaction rates \citep{Kunz02}. If the fundamental parameters of stellar remnants are accurately constrained for large and well understood samples, Galactic evolution can be derived from the space density \citep{holbergetal02-1, holbergetal08-1, giammicheleetal12-1, sionetal14-1, hollands18}, kinematic properties \citep{wegg12,anguiano17}, mass distribution \citep{bergeronetal92-1, liebertetal05-1, falconetal10-1, tremblayetal13-1, tremblayetal16-1}, and age or luminosity distributions \citep{catalanetal08-2, giammicheleetal12-1,tremblayetal14-1,rebassa-mansergasetal15-1,kilicetal17-1}.

Large, well-defined samples are also the necessary starting point in searching for rare sub-types of white dwarfs like: magnetic white dwarfs \citep{gaensickeetal02-5, schmidtetal03-1, kuelebietal09-1,kepleretal13-1, hollandsetal15-1}, pulsating stars (\citealt{castanheiraetal04-1, greissetal14-1, gentilefusilloetal16-1}), white dwarfs at the extremes of the mass distribution \citep{vennes+kawka08-1, brownetal10-1, hermesetal14-1}, stellar remnants with unresolved low mass companions \citep{farihietal05-1, girvenetal11-1, steeleetal13-1}, exotic atmospheric compositions \citep{schmidtetal99-1, dufouretal10-1,gaensickeetal10-1, kepler+koester16-1}, close double-degenerates \citep{marshetal04-1, parsonsetal11-1}, metal polluted white dwarfs \citep{sionetal90-1, zuckermanetal98-1, dufouretal07-2, koesteretal14-1, raddietal15-1} or degenerate stars with dusty or gaseous planetary debris discs \citep{gaensickeetal06-3,farihietal09-1, debesetal11-2, wilsonetal14-1,manseretal16-1}. Each one of these exotic sub-classes has extremely powerful applications in diverse areas of astronomy, from exo-planetary science to type Ia SN and cosmology.

Historic methods to identify white dwarfs include searches for UV-excess objects (e.g., the Palomar Green Survey; \citealt{Green86}, and the Hamburg/ESO survey; \citealt{hamburg}), which are restricted to the detection of blue and thus relatively hot and young white dwarfs, and the use of reduced proper motion as a proxy for their distance (e.g., \citealt{luyten79,lépine05,gentilefusilloetal15-1}), which allows the recovery of the faint end of the luminosity function. The vast majority of the $\simeq 33\,000$ spectroscopically confirmed white dwarfs known to date were discovered in the last 20 years thanks to large area spectroscopic surveys, most notably the Sloan Digital Sky Survey (SDSS; \citealt{yorketal00-1},  \citealt{eisensteinetal06-1}, \citealt{Kleinmanetal13-1}, \citealt{kepleretal16-2}). While this sample has been of fundamental importance for many white dwarf population studies to date, it suffers from severe selection biases, is largely incomplete, and is dominated by relatively hot ($T_{\rm eff} >$ 10\,000\,K) and young stars (cooling age $<$ 1\,Gyr). Furthermore, the full extent of the selection effects in the SDSS white dwarfs (non-static observing strategy, colour bias, magnitude limits, etc) are very difficult to quantify \citep{degennaro08,gentilefusilloetal15-1,tremblayetal16-1}, as the vast majority of these objects were serendipitous discoveries.
Consequently, the southern hemisphere remains a largely unexplored territory with $\lesssim$ 15 per cent of the white dwarfs known prior to {\it Gaia} located below the celestial equator. While large catalogues of white dwarf candidates based on colours and reduced proper motion compiled by, e.g., \citealt{harrisetal03-1}, \citealt{gentilefusilloetal15-1}, \citealt{munn17}, and \citealt{gentilefusilloetal17-1} circumvent many of the biases of the spectroscopic samples, they are still limited by the availability of deep multi-band photometry and accurate proper motions.

The European Space Agency (ESA) astrometric mission {\it Gaia} is the successor of the Hipparcos mission. {\it Gaia}  determined positions, parallaxes, and proper motions for $\approx1$  per cent of the stars in the Galaxy and it aims to be complete across the full sky down to {\it Gaia} $G = 20-21$ magnitudes \citep{perrymanetal01-1,Gaiaetal16-1}.  The final data release is expected to have a parallax precision better than 10 per cent for 95 per cent of the white dwarfs \citep{torresetal05-1,carrascoetal14-1}.

{\it Gaia} Data Release~1 only included six directly detected degenerate stars \citep{tremblayetal17-1}. By contrast, {\it Gaia} Data Release~2 (DR2; \citealt{gaiaDR2-ArXiV-1}) is more complete by orders of magnitude, and it provides precise astrometry \citep{gaiaDR2-ArXiV-2} as well as $G_{\rm BP}$ (330--680 nm), $G_{\rm RP}$ (640--1000 nm), and $G$ (330--1000 nm) passband photometry \citep{gaiaDR2-ArXiV-3}. 
We note that {\it Gaia} low-resolution spectrophotometry is not yet available in DR2. Furthermore, Radial Velocity Spectrometer (RVS) measurements in the region of the Ca triplet around 860 nm \citep{gaiaDR2-ArXiV-4} are of little relevance for white dwarfs as most of them are featureless in this region or too faint. 

{\it Gaia} DR2 allows for the first time the identification of field white dwarfs in an absolute magnitude versus colour (Hertzsprung-Russell, H-R) diagram, a method that has successfully been employed in the past 20 years to identify white dwarfs in clusters \citep[see, e.g.,][]{renzini96,richer97}. This represents the greatest opportunity to identify a large catalogue of white dwarfs over the entire sky  with as little colour and proper motion bias as possible. 

Following on from our overview of {\it Gaia} stellar remnants in the local 20\,pc sample \citep{hollands18}, we present a catalogue of $\simeq260\,000$ high-confidence white dwarf candidates selected from {\it Gaia} DR2 based on their {\it Gaia} parallaxes and photometry. This catalogue is meant to include all single and double {\it Gaia} white dwarfs that have a $\bp-\rp$ colour and a reliable parallax in DR2. Our catalogue includes a number of unresolved white dwarf plus main-sequence binaries as well as extremely-low-mass (ELM) white dwarfs but it is not in any way complete for these stellar types. We compare this {\it Gaia} catalogue with a new, carefully constructed sample of SDSS white dwarf candidates and assess the robustness of our {\it Gaia} selection in terms of the sky completeness of the resulting magnitude-limited sample. 

\section{White dwarf selection}

\label{selection}

\begin{table}
\centering
\caption{\label{summary} Summary of the white dwarf candidate selection in \textit{Gaia} DR2.}
\begin{tabular}{lr}
\hline\\[-1.5ex]
Total number of sources in \textit{Gaia} DR2 & 1\,692\,919\,135\\
Sources in initial colour-\Gabs~cuts (Eqs.\,1-5) & 8\,144\,735\\
Objects after quality filtering (Eqs.\,6-7) & 960\,845\\
Galactic plane objects removed (Eqs.\,8-11)& 127\,529\\
Magellanic clouds objects removed (Eqs.\,12-13)& 346\,675\\
Final size of catalogue & \finsize \\
High-confidence candidates ($\Pwd>0.75$) & 262\,480\\
\hspace{0.5cm}of which with $G\leq16$ & 1952\\
\hspace{0.5cm}of which with $16<G\leq18$ & 19\,648\\
\hspace{0.5cm}of which with $18<G\leq20$ & 158\,483\\
\hspace{0.5cm}of which with $G>20$ & 82\,397\\

\hline
\end{tabular}
\end{table}

\begin{figure}
\includegraphics[width=\columnwidth]{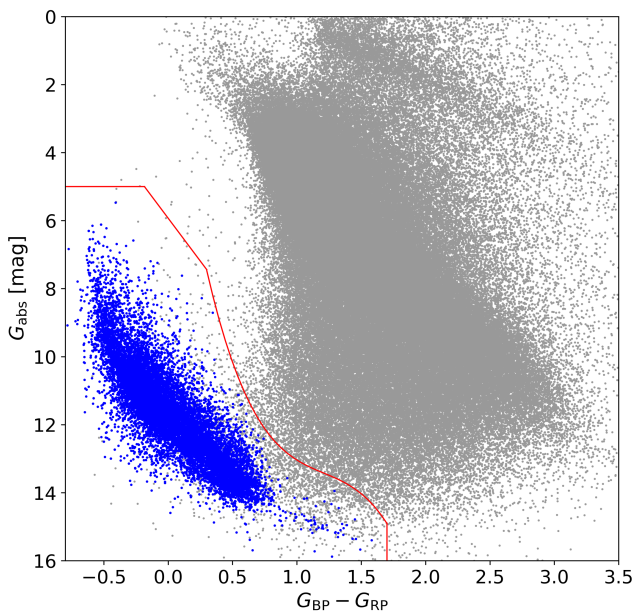}
\caption{\label{initial_cut} \textit{Gaia} H-R diagram showing a representative sample of objects (selected randomly using the $\textsc{random\_index}$ {\it Gaia} parameter) with $\textsc{parallax\_over\_error} >1$ (gray points). Spectroscopically confirmed white dwarfs (from \citealt{gentilefusilloetal15-1, hollandsetal17-1}) used to broadly define the white dwarf locus are over-plotted in blue. The initial cuts adopted for our selection are shown as red solid lines.}
\end{figure}

In order to assess the total parameter space spanned by stellar remnants in the \textit{Gaia} H-R diagram, we began by retrieving all available \textit{Gaia} data for the spectroscopically confirmed SDSS DR10  white dwarfs contained in the catalogue of \citet{gentilefusilloetal15-1}. In order to also sample the locus populated by cool white dwarfs (\Teff\ $<$ 6000\,K) we also include the SDSS objects identified by \citet{hollandsetal17-1}.  We used  this sample to define a set of broad cuts in the H-R diagram which contain  the entire parameter space spanned by white dwarfs, while attempting to exclude the area dominated by the main sequence  (Fig.\,\ref{initial_cut}). At this stage we focused on including all confirmed white dwarfs and no real effort was invested in excluding contaminant objects.

\begin{align}
\textsc{parallax\_over\_error} &>1
\end{align}
\begin{align}
\textsc{and~}\Gabs & > 5\\
\textsc{and~}\Gabs & > 5.93 + 5.047 \times (\bp-\rp) \\
\textsc{and~}\Gabs &> 6\times(\bp-\rp)^{3} \nonumber\\
& - 21.77\times(\bp-\rp)^{2} + \nonumber\\
& + 27.91\times(\bp-\rp) + 0.897 \\
\textsc{and~} (\bp-\rp) &< 1.7~,
\end{align}
{\noindent}where \Gabs\ is defined as:\\ \textsc{phot\_g\_mean\_mag}$+5\times(\log_{10}(\textsc{parallax}/1000)+1)$. Although our selection is based on the distribution of sources in the H-R diagram, we do neither make use of distances measured by \textit{Gaia} parallaxes nor make use of any Galactic simulations which would require a careful treatment for objects with parallax uncertainties greater than 10 per cent \citep{lurietal18-1}.
Eq. 1$-$5 include 8\,144\,735 \textit{Gaia} sources and serve primarily to limit the space within which to carry out further selections (Table\,\ref{summary}). The resulting \textit{Gaia}\,DR2 sample contains large numbers of objects with bad astrometry and/or photometry, and unreliable detections. Hence, an additional refined selection within this sample is necessary to establish a clean catalogue of white dwarf candidates.
We cross-matched the positions of all objects within our initial cuts with all 4\,851\,200 spectra currently available within SDSS\,DR14. 
We found a total of $\simeq 36\,000$ objects with spectra and proceeded to separate white dwarfs and contaminants  by visual inspection. The contaminant objects are dominated by main sequence stars and subdwarfs, but a small number of quasars also clear our initial selection. This \textit{Gaia}-SDSS spectroscopic sample is described in more details in Section\,\ref{spec_SDSS_sect}.
The spectroscopically confirmed white dwarfs and contaminants can be used to test the effect that any additional quality filtering will have on the completeness of our white dwarf selection.
Unless otherwise stated, we use for our filtering the measurements and flags provided in DR2. 

\begin{figure*}
\includegraphics[width=2\columnwidth]{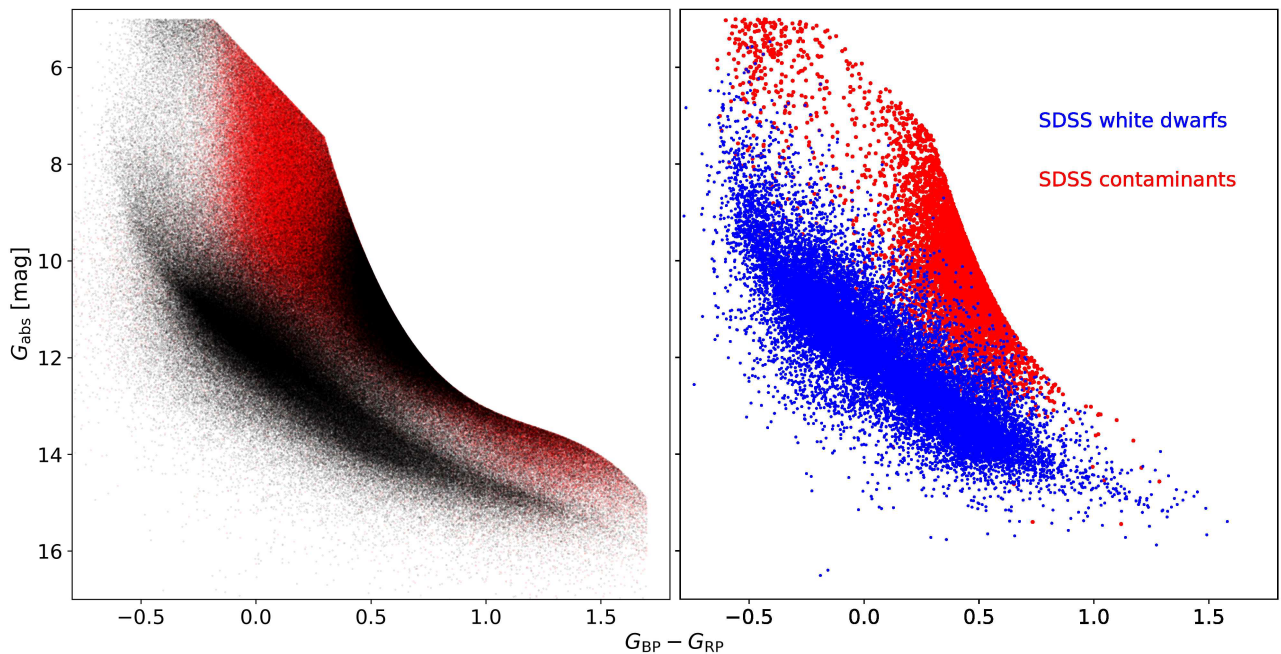}
\caption{\label{combined} \textit{Left panel:} \textit{Gaia} H-R diagram of all 960\,845 objects selected using our absolute magnitude and colour cuts, and quality filtering (Eqs.\,1-7). The 127\,529 Galactic plane and 346\,675 Magellanic cloud objects removed by our additional filtering (Eqs.\,8-13) are shown in red. \textit{Right panel:} Distribution of spectroscopically confirmed SDSS white dwarfs (blue) and contaminants (red) included in our final \textit{Gaia} sample.}
\end{figure*}

The flags with the largest impact on our selection are now described in turn. The value of $\textsc{phot\_bp\_rp\_excess\_factor}$ $([f_{\rm BP} + f_{\rm RP}]/f_{\rm G}$, where $f$ is the observed flux in e-/s) indicates whether the three \textit{Gaia} photometric bands are consistent with the assumption of an isolated source and can be used to identify objects with unreliable colours or a bright sky background \citep[see Section 8 of][]{gaiaDR2-ArXiV-3}. $\textsc{astrometric\_excess\_noise}$ is a measure of the residuals in the astrometric solution for the source, and can therefore be used to identify objects with unreliable parallax measurements \citep{gaiaDR2-ArXiV-2}. $\textsc{astrometric\_sigma5d\_max}$ is a five-dimensional equivalent to the semi-major axis of the \textit{Gaia} position error ellipse and is useful for filtering out cases where one of the five parameters, or some linear combination of several parameters, is particularly bad \citep{gaiaDR2-ArXiV-2}. The quality cuts in $\textsc{astrometric\_excess\_noise}$ and $\textsc{phot\_bp\_rp\_excess\_factor}$ proposed in \citet{gaiaDR2-ArXiV-5} do indeed provide a very clean sample of objects, but they also exclude over 15 per cent of the known SDSS white dwarfs brighter than $G=20$. Striving to construct a sample as complete as possible we defined the following set of quality cuts based on \textit{Gaia} flags and measurements, which exclude non-white dwarf contaminants and objects with poor measurements, while preserving most of the degenerate stars. 
\begin{align}
\textsc{phot\_bp\_rp\_excess\_factor} &< (1.7 +  0.06 \nonumber\\ 
& \times \textsc{bp\_rp}^{2})\\
\textsc{and}\ (\textsc{astrometric\_sigma5d\_max}&<1.5 ~ \textsc{or}\\
(\textsc{astrometric\_excess\_noise}&<1 \nonumber\\
\textsc{and}~ \textsc{parallax\_over\_error}&>4 \nonumber\\
\textsc{and}~\textsc{sqrt}(\textsc{pmra}^{2}+\textsc{pmdec}^{2})&>10\,mas))\nonumber
\end{align}
Eq.\,7 is designed to exclude objects flagged to have sub-optimal five-parameter solution, without however rejecting objects which still have reliable parallaxes (low \textsc{astrometric\_excess\_noise} and at least 4$\sigma$ parallax measurements) and significant proper motions.
These quality cuts exclude only $<2$ per cent of the known SDSS white dwarfs with $G<20$, while removing 25\,per\,cent of the similarly bright  SDSS contaminants. Using this selection also eliminates the vast majority of \textit{Gaia} objects with  poor measurements, bringing the sample size to 960\,845 (Fig.\,\ref{combined}). Beyond $G>20$, however, the quality of \textit{Gaia} photometric and astrometric data quickly deteriorates and only 60 per cent of the known SDSS white dwarfs with $G>20$ are retrieved by our quality cuts. \textit{Gaia} measurements get significantly worse in areas of the sky with high stellar density and main-sequence stars located in the Galactic plane have such large scatter in the H-R diagram that they contaminate most of the white dwarf locus. In order to filter these sources we calculated a $\textsc{density}$ parameter for all 960\,845 objects included by Eqs.\,1-7. This was done by dividing up the sky into approximately $10'\times10'$ tiles, with sides defined by lines of constant RA and Dec. The number of \textit{Gaia} DR2 targets in each tile was calculated and converted to a density of sources per square degree. Then each catalogue object was assigned the density corresponding to the tile that it fell into.
We then proceeded to apply stricter quality filters on objects closer to the Galactic plane as
\begin{align}
|b| &<25^{\circ}\\
\textsc{and~}\textsc{density} &>100\,000\,\mathrm{deg}^{-2}\\
\textsc{and~}\textsc{phot\_bp\_rp\_excess\_factor} &> (1.0 +0.015\\
& \times \textsc{bp\_rp}^{2}) \nonumber\\
\textsc{and~}\textsc{phot\_bp\_rp\_excess\_factor} &< (1.3 +0.06\\
& \times \textsc{bp\_rp}^{2}) \nonumber
\end{align}
The threshold \textsc{density} value of 100\,000\,deg$^{-2}$ (Eq.\,9) was determined by attempting to eliminate Galactic plane sources with the highest scatter in $\bp-\rp$.
This additional filtering removes 127\,529 objects from the sample. Visual inspection of the distribution in H-R space of our sample of SDSS spectroscopic objects and of the remaining 833\,316 \textit{Gaia} objects (Table\,\ref{summary}), reveals an over-density of \textit{Gaia} sources in areas that are scarcely populated by SDSS targets. These over-abundant \textit{Gaia} objects are almost exclusively located in  the Magellanic clouds. 
This over-density is therefore spurious and we need to remove extra-galactic sources from our sample as efficiently as possible while attempting to preserve foreground stars.
We select two broad areas which encompass the Magellanic clouds defined as two rectangles, the first one centred on $\alpha= 22.5^{\circ}$ $\delta= -75.0^{\circ}$ extends $45^{\circ}$ in right ascension  and $30^{\circ}$ in declination; the second one centred on  $\alpha=82^{\circ}$ $\delta= -68.0^{\circ}$ extends $55^{\circ}$ in right ascension and $35^{\circ}$ in declination. 
Within this space we adopt the following further filtering on objects in crowded areas:
\begin{align}
\textsc{density} &>11\,000\,\mathrm{deg}^{-2}\addtocontents{file}{text}\\
\textsc{and~}\textsc{parallax\_over\_error} &>10
\end{align}
As we intend to completely remove the over-density of objects due to the Magellanic clouds, the adopted \textsc{density} threshold (Eq.\,12) is the median \textsc{density} value over the entire sky outside of the plane and the Magellanic clouds. This final filtering brings the size of our \textit{Gaia} sample to the final value of \finsize~ (Fig.\,\ref{combined}, Table\,\ref{summary}).

\begin{figure}
\includegraphics[width=\columnwidth]{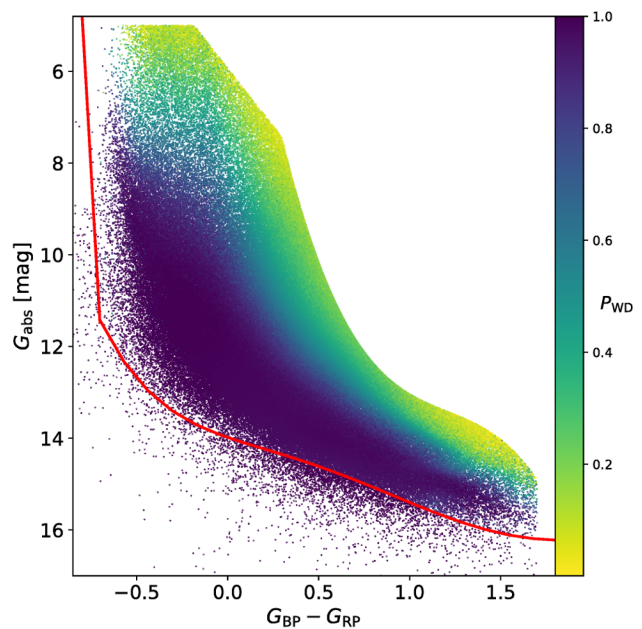}
\caption{\label{pwd_gradient} \textit{Gaia} H-R diagram of all \finsize\ objects in our catalogue. The colour scale indicates the \Pwd\ of each object. All objects below and to the left of the solid red lines are assigned a \Pwd \_\textsc{flag}.}
\end{figure}

Fig.\,\ref{combined} shows that, even after applying all the filtering on \textit{Gaia} parameters described above, there  remains an overlap between white dwarfs and other stars. Consequently selecting white dwarfs with any cut in H-R space alone would result in an incomplete, inhomogeneous, and contaminated sample. To overcome this problem, we developed a selection analogous to the one presented in \citet{gentilefusilloetal15-1}, i.e. we use the sample of spectroscopically confirmed SDSS white dwarfs and contaminants we developed, to calculate \textit{probabilities of being a white dwarf} ($P_{\rm WD}$) for all objects in our \textit{Gaia} sample. We used a total of 21\,456 spectroscopically confirmed single white dwarfs and 5982 contaminants (stars and QSOs) still included in the \textit{Gaia} sample after all the quality filtering described above. We used these objects to map the distribution of white dwarfs and contaminants in H-R ($\bp-\rp$, \Gabs) space (Fig.\,\ref{combined}).
In order to create a continuous map, every object was treated as a 2D Gaussian, the width of which reflects the uncertainty in $\bp-\rp$ vs. \Gabs\ space of the object. These Gaussians were normalised so that their volume equals unity and therefore the sum of the integrals of all the Gaussians in the map is equal to the number of objects in the training sample. This results in two continuous smeared-out density maps for white dwarfs and for contaminants.
We then defined a probability map as the ratio of the white dwarf density map to the sum of both density maps and then use this map to calculate the \Pwd\ of any given object by integrating the product of its Gaussian distribution in H-R space with the underlying probability map, giving a direct indication of how likely it is for the source to be a white dwarf. Regions outside our H-R cuts (Eqs.\,1-5) are considered to have zero probability of being a white dwarf. 
Our SDSS training sample contained only  some objects with very blue ($\bp-\rp$ $<-0.5$) colours or large absolute magnitudes ($\Gabs >15$) resulting in a patchy probability map with large areas with no information. 
These regions cover areas of the H-R diagram that we assume should be populated mostly by white dwarfs but
\Pwd\ values calculated for objects in these regions are not reliable.  We defined two polynomial lines (Eqs.\,14-15) in H-R space,
\begin{align}
\Gabs & > (\bp-\rp) \times 68.42 + 59.50\label{corr1}\\
\Gabs &> (\bp-\rp)^{5} \times 0.25 \nonumber\\
& - (\bp-\rp)^{4}\times 1.3 \nonumber\\
& + (\bp-\rp)^{3}\times 2.14 \nonumber\\
& - (\bp-\rp)^{2}\times 0.98 \nonumber\\
&+(\bp-\rp) \times 1.37 +13.98
\label{corr2}
\end{align}
below which the $\Pwd$ values should be treated with caution. In the final catalogue we include a \Pwd \_\textsc{flag}  column to indicate which objects fall below these lines (Fig.\,\ref{pwd_gradient}).

\begin{figure*}
\includegraphics[width=2\columnwidth]{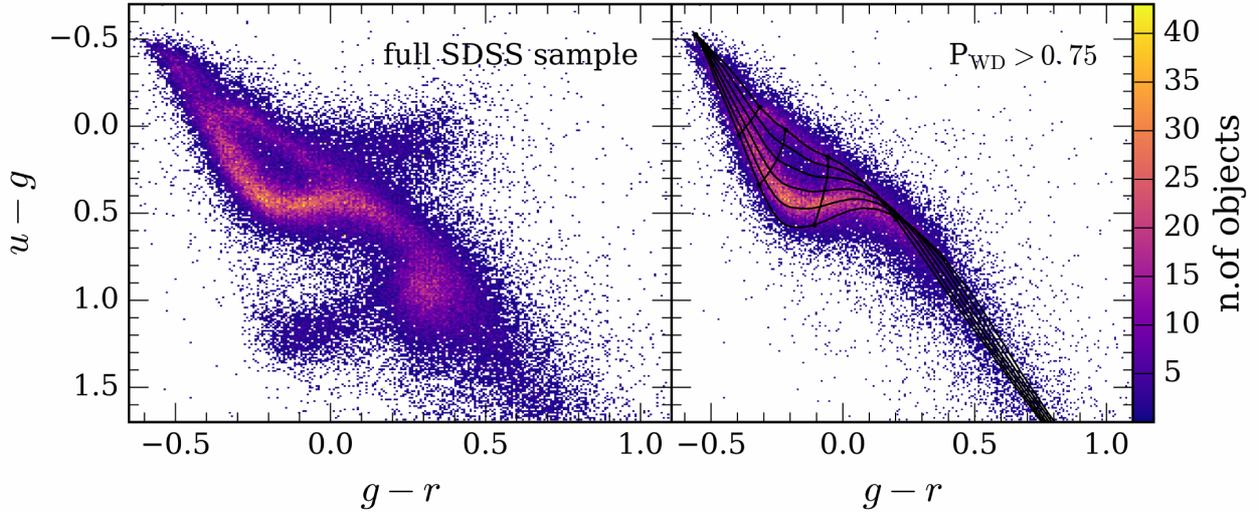}
\caption{\label{sdss_colour} \textit{Left panel:} Distribution in $u-g, g-r$ colour-colour space, of  96\,938 objects with clean SDSS photometry from our \textit{Gaia} catalogue of white dwarf candidates. \textit{Right panel:} Same distribution as right panel, but only displaying  objects with \Pwd $>$ 0.75. The distribution of these high-confidence candidates closely matches the pure-H atmosphere white dwarf cooling tracks from \citet{tremblayetal11-1} shown in black overlay. }
\end{figure*}

\begin{figure}
\includegraphics[width=0.95\columnwidth]{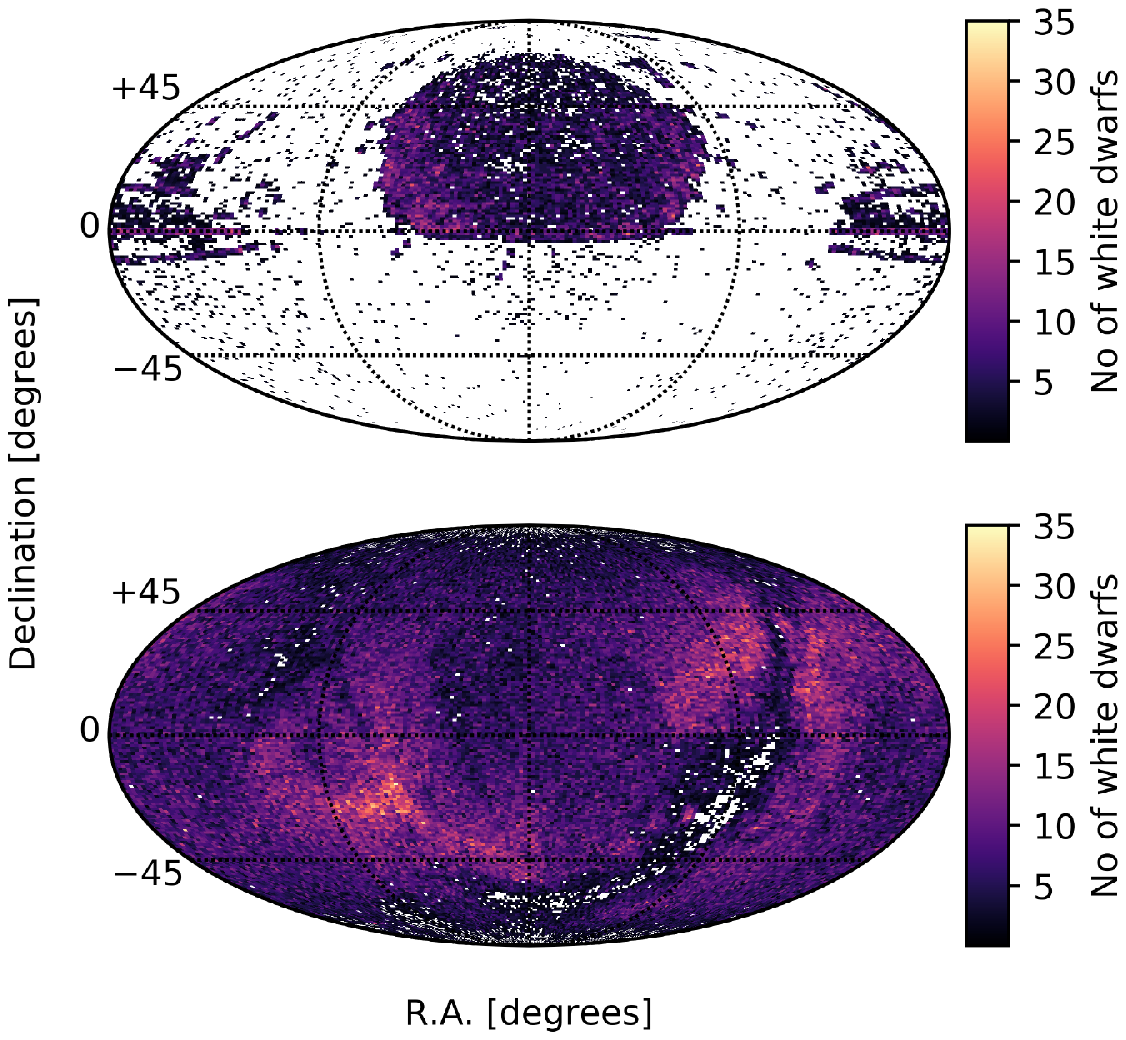}
\caption{\label{Gaia_allsky} \textit{Top panel:} Sky density of the $\simeq 33\,000$ known white dwarfs before \textit{Gaia} DR2  \citep{dufouretal17-1,kepleretal16-1,hollandsetal17-1,kilkennyetal15-1,gentilefusilloetal15-2,gentilefusilloetal17-1}. \textit{Bottom panel:} Sky density of \textit{Gaia} DR2 white dwarf candidates with $\Pwd>0.75$ from the catalogue presented in this article.}
\end{figure}

Selecting sub-samples of white dwarf candidates based on \Pwd\ allows for a flexible compromise between completeness and level of potential contamination. As a generic guideline selecting objects with $\Pwd> 0.75$ recovers 96 per cent of the spectroscopically confirmed SDSS white dwarfs in the catalogue and only 1 per cent of the contaminant objects (Fig.\,\ref{sdss_colour}). Cleaner, but significantly less complete, white dwarf subsets can be obtained by combining our \Pwd\ values with additional cuts in \textit{Gaia} quality parameters (e.g., $\textsc{astrometric\_excess\_noise}$) stricter than those already adopted in our selection. In total we estimate the final catalogue to contain $\simeq260\,000$ genuine white dwarfs, nearly an eight-fold increase in sample size compared to the number of white dwarfs known before the release of \textit{Gaia}\,DR2 (Fig.\,\ref{Gaia_allsky}). 
Furthermore only 15 per cent of the previously known white dwarfs  are located in the southern hemisphere and this new sample brings our coverage of the southern sky on level with that of the northern one \citep{gentilefusilloetal17-2}.

\section{The catalogue of white dwarfs}
\label{catalogue_section}

\begin{figure*}
\includegraphics[width=1.7\columnwidth]{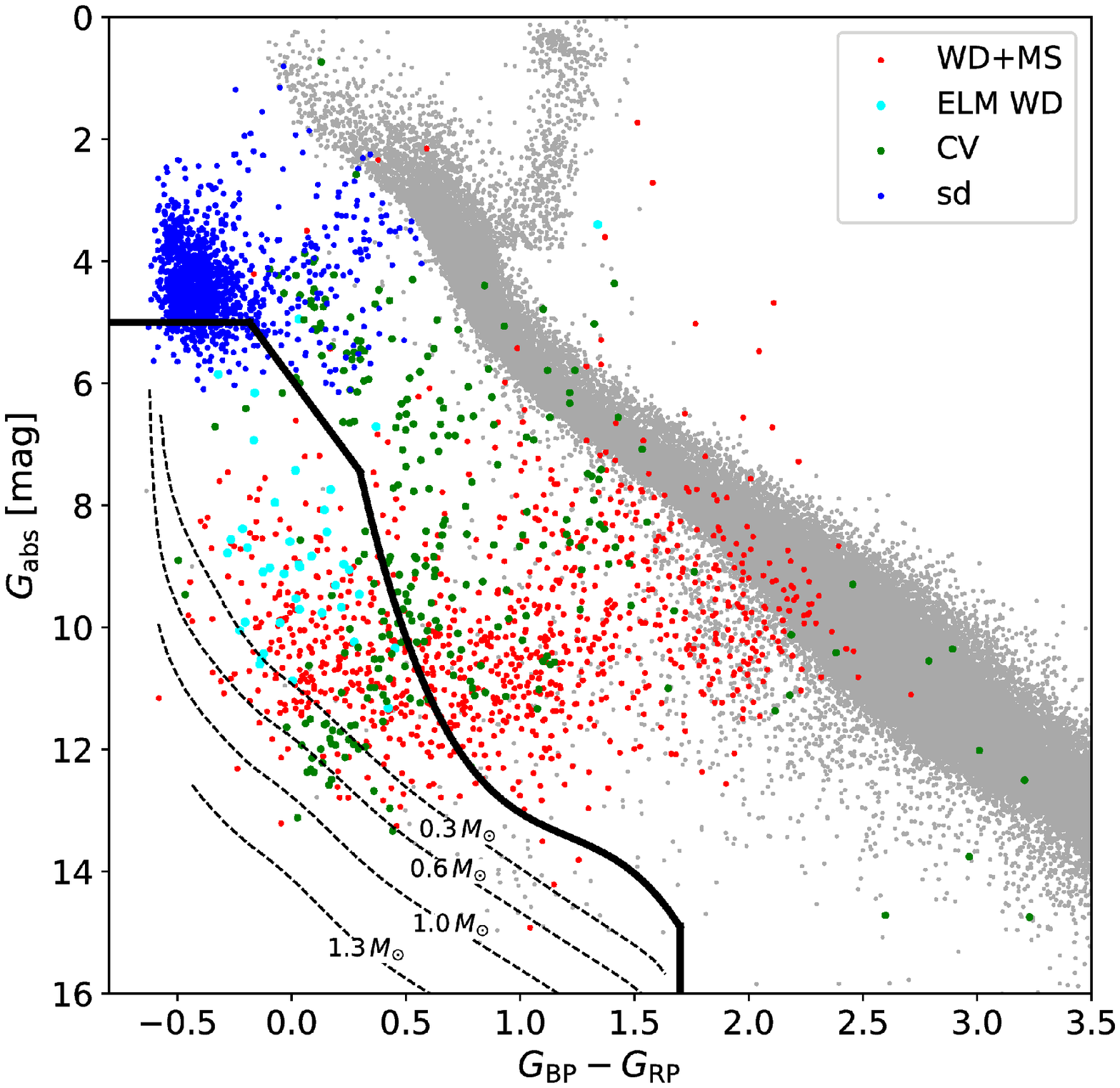}
\caption{\label{quasi_WD} H-R diagram showing the position of various families of objects closely related to single white dwarfs. A representative sample of subdwarfs from \citet{geieretal17-1} is plotted in blue. SDSS white dwarfs + main sequence binaries (WD+MS, \citealt{rebassa-mansergasetal10-1}) and cataclysmic variables (CV, \citealt{szkodyetal02-2,szkodyetal03-2,szkodyetal04-1,szkodyetal06-1,szkodyetal07-2,szkodyetal09-1,szkodyetal11-1}) are plotted in red and green, respectively. Known extremely-low-mass white dwarfs (ELM WD, \citealt{brownetal16-1}) are plotted in cyan.
An arbitrary and clean sample of \textit{Gaia} objects is also plotted in gray for reference. The dashed black lines represent cooling tracks for DA white dwarfs at different masses. The solid black line represents the initial cuts that we have used in constructing our catalogue of \textit{Gaia} white dwarf candidates.}
\end{figure*}

The selection described in Section\,\ref{selection} results in a sample of \finsize\ $Gaia$ sources for which we calculated \Pwd\ values based on the distribution of spectroscopically confirmed SDSS white dwarfs and contaminants in $G_{\rm BP}-G_{\rm RP}$ colour and \Gabs\ magnitude. We define this sample as our final catalogue of \textit{Gaia} DR2 white dwarf candidates. With this catalogue we aim to be as complete as possible in recovering single white dwarfs and double degenerate  binaries with reliable \textit{Gaia} data. However, as illustrated in Fig.\,\ref{quasi_WD} a number of subdwarfs, white dwarf plus main sequence binaries and cataclysmic variables (CVs) could also be included in our selection. We are not able to completely exclude these objects from our catalogue nor do we aim to be fully inclusive of them. While the vast majority of subdwarfs included in the final selection will have relatively low \Pwd\ values, some white dwarf main sequence binaries, CVs and ELM white dwarfs will have \Pwd\ values comparable to those of typical single stellar remnants.

The format of the catalogue is described in detail in Table\,\ref{Table format}.
Column $1$ of the catalogue contains the WD\,J name we assigned to the objects following the proposed convention described in Section\,\ref{new_name}.
Columns 3-28 are directly acquired from the DR2 {\sc gaia\_source} table \citep{gaiaDR2-ArXiV-1}. 
We also include the SDSS name and photometry of all $Gaia$ white dwarfs with a reliable SDSS match. We find that the \textsc{SDSSDR}9\textsc{\_best\_neighbour} table provided in \textit{Gaia} DR2 does not include 2375 SDSS objects which have reliable matches in \textit{Gaia}. We were not able to determine any specific selection effect which caused these objects to be excluded. Therefore we performed our own cross match between \textit{Gaia} and SDSS using a matching radius of two arcseconds and accounting for the difference in epoch of observation and the proper motion of each object. 
Some SDSS photometric objects which appear in SDSS DR7 were not included again in subsequent data releases \citep{gentilefusilloetal15-1}, so our cross-match with SDSS was carried out both on DR7 and DR14.

The entries in the remaining columns (43-56) result from our model fits to the $Gaia$ data for a subset of the sample and are fully described in Section\,\ref{sect_atm_par}.

The \textit{Gaia} catalogue does not include the error on magnitude but this can be calculated from the relative error on the electron flux per second as

\begin{align}
\sigma(G^S) = \frac{2.5}{\ln(10)}\frac{\sigma(f^S)}{f^S}
\end{align}

{\noindent}where $S$ refers to any {\it Gaia} passband, $G$ is the magnitude, and $f$ is the flux.

\subsection{A new white dwarf naming convention}
\label{new_name}

As in many areas of astronomy, individual white dwarfs are often known by more than one name, e.g. vMa2 = Wolf\,28 = EGGR\,5 = WD\,0046+051 or GD362 = G\,204--14 = NLTT\,44986 = EGGR\,545 = WD\,1729+371, which is often source of confusion. \citet{mccook+sion87-1} introduced the ``WD number'' as a unifying identifier, which is composed of the first four digits of right ascension (hours and minutes), the sign of the declination, the first two digits of the declination (degrees) and a third digit which expresses the minutes of the declination as a truncated fraction of a degree, where the coordinates are in the 1950 equinox. Over the last two decades, the number of known white dwarfs has rapidly grown from just over 2\,000 \citep{mccook+sion99-1} to well over 30\,000 \citep{eisensteinetal06-1, Kleinmanetal13-1, gentilefusilloetal15-1, kepleretal15-1, kepleretal16-1, gentilefusilloetal17-2, raddietal17-1}, and it is clear that with the next order-of-magnitude increase in the sample size presented here, the historically used naming convention will no longer be suitable. We therefore propose to adopt a new naming convention, which shall account for proper motions. WD\,JHHMMSS.SS$\pm$DDMMSS.S will be defined as the white dwarf coordinates in IRCS, at equinox 2000 and epoch 2000. This definition should be sufficient to avoid duplicate names even in the era of LSST (estimated to identify over 13 million white dwarfs; \citealt{LSST09-1}), except in the densest environments, such as globular clusters. We have determined the white dwarf names in Table\,\ref{Table format} following this convention, using the \textit{Gaia} IRCS coordinates and proper motions, which are provided at equinox and epoch 2015.5.

\begin{table*}
\centering
\caption{\label{Table format} Format of the \textit{Gaia} DR2 Catalogue of white dwarfs. The full catalogue can be accessed online via the VizieR catalogue access tool.}
\begin{tabular}{lll}
\hline
\hline
Column & Heading & Description\\
\hline
1 & \textsc{White\_dwarf\_name} & WD\,J + J2000 ra (hh mm ss.ss) + dec (dd mm ss.s), equinox and epoch 2000\\
2 & \textsc{\Pwd} & The \textit{probability of being a white dwarf} (see Sect.\,\ref{selection})\\ 
3 & \textsc{Designation} & Unique Gaia source designation (unique across all Data Releases)\\
4 & \textsc{source\_id} & Unique Gaia source identifier (unique within a particular Data Release)\\
5 & \textsc{ra} & Right ascension (J2015.5) [deg]\\
6 & \textsc{ra\_error} & Standard error of right ascension ($\times cos(\delta$)) [mas]\\ 
7 & \textsc{dec} & Declination (J2015.5) [deg]\\
8 & \textsc{dec\_error} & Standard error of declination [mas]\\ 
9 & \textsc{parallax} & Absolute stellar parallax of the source at J2015.5 [mas]\\
10 &\textsc{parallax\_error} & Standard error of parallax [mas]\\
11& \textsc{pmra} & Proper motion in right ascension  ($\times cos(\delta$)) [mas/yr]\\
12& \textsc{pmra\_err} & Standard error of proper motion in right ascension [mas/yr]\\
13& \textsc{pmdec} & Proper motion in right declination [mas/yr]\\
14& \textsc{pmdec\_err} & Standard error of proper motion in right declination [mas/yr]\\
15& \textsc{astrometric\_excess\_noise} & Measure of the residuals in the astrometric solution for the source [mas] (see Sect.\,\ref{selection})\\
16 & \textsc{astrometric\_sigma5d\_max} & Five-dimensional equivalent to the semi-major axis of the Gaia position error ellipse\\ 
& & [mas] (see Sect.\,\ref{selection})\\
17 & \textsc{phot\_$G$\_mean\_flux} & Gaia $G$-band mean flux [e-/s]\\
18 & \textsc{phot\_$G$\_mean\_flux\_error} &  Error on $G$-band mean flux [e-/s]\\
19& \textsc{phot\_$G$\_mean\_mag} & Gaia $G$-band mean magnitude (Vega scale) [mag]\\
20 & \textsc{phot\_$G_{\rm BP}$\_mean\_flux} & Integrated $G_{\rm BP}$ mean flux [e-/s]\\
21 & \textsc{phot\_$G_{\rm BP}$\_mean\_flux\_error} &  Error on integrated $G_{\rm BP}$ mean flux [e-/s]\\
22& \textsc{phot\_$G_{\rm BP}$\_mean\_mag} & Integrated $G_{\rm BP}$ mean magnitude (Vega scale) [mag]\\
23 & \textsc{phot\_$G_{\rm RP}$\_mean\_flux} & Integrated $G_{\rm RP}$ mean flux [e-/s]\\
24 & \textsc{phot\_$G_{\rm RP}$\_mean\_flux\_error} &  Error on integrated $G_{\rm RP}$ mean flux [e-/s]\\
25 & \textsc{phot\_$G_{\rm RP}$\_mean\_mag} & Integrated $G_{\rm RP}$ mean magnitude (Vega scale) [mag]\\
26 & \textsc{phot\_$G_{\rm BP}$\_$G_{\rm RP}$\_excess\_factor} & $G_{\rm BP}$/$G_{\rm RP}$ excess factor estimated from the comparison of the sum of integrated\\ 
& & $G_{\rm BP}$ and $G_{\rm RP}$ fluxes with respect to the flux in the $G$ band (See Sect.\,\ref{selection})\\
27  & l & Galactic longitude [deg]\\
28 & b & Galactic latitude [deg]\\
29 & \textsc{Density} & The number of Gaia sources per square degree around this object (see Sect.\,\ref{selection}).\\
30 & \textsc{\Pwd\_flag} & If 1 it indicates the \Pwd\ value could be unreliable (see Sect.\,\ref{selection}, Fig.\,\ref{pwd_gradient})\\
31 & \textsc{$A_G$} & Extinction [mag] in the Gaia $G$ band derived from E($B-V$) values from\\ 
& & \citet{schlafy} (see Sect.\,\ref{sect_atm_par})\\
32 & \textsc{SDSS\_name} & SDSS object name if available (SDSS + J2000 coordinates) \\
33 & \textsc{$u$mag} & SDSS $u$ band magnitude [mag]\\
34 & \textsc{$u$mag\_err} & SDSS $u$ band magnitude uncertainty [mag]\\
35 & \textsc{$g$mag} & SDSS $g$ band magnitude [mag]\\
36 & \textsc{$g$mag\_err} & SDSS $g$ band magnitude uncertainty [mag]\\
37 & \textsc{$r$mag} & SDSS $r$ band magnitude [mag]\\
38 & \textsc{$r$mag\_err} & SDSS $r$ band magnitude uncertainty [mag]\\
39 & \textsc{$i$mag} & SDSS $i$ band magnitude [mag]\\
40 & \textsc{$i$mag\_err} & SDSS $i$ band magnitude uncertainty [mag]\\
41 & \textsc{$z$mag} & SDSS $z$ band magnitude [mag]\\
42 & \textsc{$z$mag\_err} & SDSS $z$ band magnitude uncertainty\\
43 & \textsc{\Teff\_(H)} & Effective temperature [K] from fitting the dereddened $G$, $G_{\rm BP}$, and $G_{\rm RP}$ absolute fluxes\\
& & with pure-H model atmospheres (see Sect.\,\ref{sect_atm_par})\\
44 & \textsc{$\sigma$\_\Teff\_(H)} & Uncertainty on \Teff~[K]\\
45 & \textsc{$\log$\_$g$\_(H)}& Surface gravity [cm/s$^2$] from fitting the dereddened $G$, $G_{\rm BP}$, and $G_{\rm RP}$ absolute fluxes\\
& & with pure-H model atmospheres (see Sect.\,\ref{sect_atm_par})\\
46 & \textsc{$\sigma$\_$\log\_g$\_(H)} & Uncertainty on $\log g$ [cm/s$^2$]\\
47 & \textsc{$M$\_(WD, H)}& Stellar mass [$\mathrm{M}_{\odot}$] resulting from the adopted mass-radius relation\\
& & and best fit parameters in columns 43-46 (see Sect.\,\ref{sect_atm_par})\\
48 & \textsc{$\sigma$\_$M$\_(WD, H)} & Uncertainty on the mass [$\mathrm{M}_{\odot}$]\\
49 & \textsc{$\chi^2$\_(H)} & $\chi^2$ value of the fit (pure-H)\\
50 & \textsc{\Teff\_(He)} & Effective temperature [K] from fitting the dereddened $G$, $G_{\rm BP}$, and $G_{\rm RP}$ absolute fluxes\\
& & with pure-He model atmospheres (see Sect.\,\ref{sect_atm_par})\\
51 & \textsc{$\sigma$\_\Teff\_(He)} & Uncertainty on \Teff [K]\\
52 & \textsc{$\log$\_$g$\_(He)}& Surface gravity [cm/s$^2$] from fitting the dereddened $G$, $G_{\rm BP}$, and $G_{\rm RP}$ absolute fluxes\\
& & with pure-He model atmospheres (see Sect.\,\ref{sect_atm_par})\\
53 & \textsc{$\sigma$\_$\log\_g$\_(He)} & Uncertainty on $\log g$ [cm/s$^2$]\\
54 & \textsc{$M$\_(WD, He)}& Stellar mass [$\mathrm{M}_{\odot}$] resulting from the adopted mass-radius relation\\
& & and best fit parameters in columns 50-53 (see Sect.\,\ref{sect_atm_par})\\
55 & \textsc{$\sigma$\_$M$\_(WD, He)} & Uncertainty on the mass [$\mathrm{M}_{\odot}$]\\
56 & \textsc{$\chi^2$\_(He)} & $\chi^2$ value of the fit (pure-He)\\
\hline
\end{tabular}
\end{table*}

\section{Atmospheric Parameters}
\label{sect_atm_par}

Only a small fraction of the white dwarfs in our catalogue have available spectroscopy and we must therefore use another technique to characterise the atmospheric parameters ($T_{\rm eff}$ and $\log g$) of the sample as a whole. Photometric surveys, such as Pan-STARRS, 2MASS, SDSS, and GALEX, only offer a partial coverage, either in magnitude range or sky area. The advantage of adding near-ultraviolet, near-infrared, or narrow-band photometry to the {\it Gaia} data set is also limited by the fact that we do not know the atmospheric compositions. Therefore, fitting additional data sets with either pure-H or pure-He model atmospheres may not necessarily improve the accuracy of the atmospheric parameters. As a consequence, we use {\it Gaia} DR2 data only to determine the atmospheric parameters, and we compare our results to Pan-STARRS and SDSS photometry for bright DA white dwarfs to understand possible systematic effects in Section\,\ref{sect_atm_par.1}. 

There is a degeneracy between $T_{\rm eff}$ and reddening when using {\it Gaia} data alone. The first step of our photometric analysis is therefore to derive an estimate for the amount of reddening. We have queried the \citet{schlegeletal98-1} reddening maps and incorporated the correction proposed by \citet{schlafy}. We have assumed that the extinction coefficient $A_{\rm G}$ in the {\it Gaia} $G$ passband scales as $0.835 \times A_{\rm V}$ based on the nominal wavelengths of the respective filters and the reddening versus wavelength dependence employed by \citet{schlafy}. The resulting $A_{\rm G}$ values are given in column 31 of our catalogue (Table\,\ref{Table format}). The reddening as a function of distance is parameterised assuming that the absorbing material along the line of sight is concentrated along the plane of the Galactic disc with a scale height of 200\,pc. Under this assumption the dereddened magnitudes are given by
\begin{equation}
G_{\rm star} = G_{\rm obs}-A_{\rm G}(1-\exp\Big(-\frac{\sin(|b|)}{200\varpi}\Big))
\end{equation}
\begin{equation}
G_{\rm BP, star} = G_{\rm BP, obs}-1.364 \times A_{\rm G}(1-\exp\Big(-\frac{\sin(|b|)}{200\varpi}\Big))
\end{equation}
\begin{equation}
G_{\rm RP, star} = G_{\rm RP, obs}-0.778 \times A_{\rm G}(1-\exp\Big(-\frac{\sin(|b|)}{200\varpi}\Big))
\end{equation}
{\noindent}where $b$ is the vertical Galactic coordinate and the parallax $\varpi$ is in arcsec. This parameterisation is slightly different to the one used in \citet{harrisetal06-1}, \citet{tremblayetal11-1}, and \citet{beaulieu14}, where interstellar absorption was assumed to be negligible within 100\,pc and to vary linearly from zero to a maximum line of sight value for $\sin(|b|)/\varpi$ between 100 and 250\,pc. We have verified that our new parameterisation provides a slightly better empirical agreement between the observed hot white dwarf cooling sequences ($G_{\rm BP}-G_{\rm RP} < 0$) for distances in the ranges $<$ 75 and 75--250\,pc, respectively. In this colour range the {\it Gaia} sample is expected to be fairly complete up to 250\,pc and therefore the properties of the dereddened white dwarfs should not depend on the distance. From this experiment we could rule out a gas scale height that is either two times larger or two times smaller than 200\,pc.

Improved {\it Gaia} DR2 reddening maps in three dimensions will eventually supersede our simple parameterisation but it is currently outside the scope of this work.

We have employed the {\it Gaia} DR2 revised quantum efficiency $S(\lambda)$ 
  for the $G$, $G_{\rm BP}$ and $G_{\rm RP}$ passbands \citep{gaiaDR2-ArXiV-3} to calculate synthetic absolute magnitudes using the relation
\begin{equation}
  M^{S} = -2.5\log\left( \frac{\int S(\lambda)F(\lambda)\lambda d\lambda}{\int S(\lambda)\lambda d\lambda} \frac{1}{(10\,{\rm pc})^2} \right)+C^S~,
  \label{eqZeroPoint1}
\end{equation}
{\noindent}where 10\,pc is expressed in cm (1\,pc = $3.08568\times10^{18}$ cm), $C^S$ is the zero point, and $F(\lambda)$ is the integrated stellar flux in erg s$^{-1}$ \AA$^{-1}$ relating to the emergent monochromatic Eddington flux $H_\lambda$ as
\begin{equation}
  F(\lambda) = 4 \pi R^2 H_\lambda(T_{\rm eff},\log g) ~,
  \label{eqZeroPoint2}
\end{equation}
{\noindent}where $R$ is the white dwarf radius. We employ standard H-atmosphere spectral models \citep{tremblayetal11-1} including the L$\alpha$ red wing absorption of \citet{kowalski06} and covering the range $1500 < T_{\rm eff}$ (K) $< 140\,000$ and $6.5 < \log g < 9.5$. For $M_{\rm WD} > 0.46\,\mathrm{M}_{\rm \odot}$, we use the evolutionary sequences with thick hydrogen layers ($M_{\rm H}/M_{\rm WD} = 10^{-4})$ of \citet[][$T_{\rm eff} \leq 30\,000$\,K, C/O-core 50/50 by mass fraction mixed uniformly]{fontaine01} and \citet[][$T_{\rm eff} >$ 30\,000\,K, pure C-core]{wood95}. For lower masses, we use the He-core cooling sequences of \citet{althaus01}. We have also computed synthetic magnitudes for He-atmosphere models \citep{bergeronetal11-1} using a mass-radius relation for thin hydrogen layers \citep[][$M_{\rm H}/M_{\rm WD} = 10^{-10}$]{fontaine01}. For the {\it Gaia} passbands Vega has a magnitude of +0.03 \citep{jordi10}, and the zero points defined with this reference are given in Table\,\ref{zeropoints} along with nominal wavelengths. The values for the pre-launch nominal {\it Gaia} filters are also given \citep{jordi10,carrascoetal14-1,tremblayetal17-1}.

\begin{table}
\centering
\caption{{\it Gaia} photometric zero points}
\begin{tabular}{llllll}
\hline
\hline
Filter & DR2 & DR2 & DR1 & DR1\\
 & $<\lambda>$ & Zero point & $<\lambda>$ & Zero point\\
\hline
$G$ & 6773.70 & $-$21.48270 &  6735.72  & $-$21.48058 \\
$G_{\rm BP}$ & 5278.58 & $-$20.95873  & 5320.63 &  $-$20.94187 \\
$G_{\rm RP}$ & 7919.08 & $-$22.20075 & 7993.39 &  $-$22.24105 \\
$G_{\rm RVS}$ & 8597.40 & $-$22.59931  & 8597.40 &  $-$22.59931 \\
\hline\label{zeropoints}
\end{tabular}
\end{table}

The dereddened observed {\it Gaia} flux $f^{\rm S}$ in the passband $S$ in units of erg cm$^{-2}$ s$^{-1}$ can be computed from the dereddened apparent {\it Gaia} magnitude in the same passband as
\begin{equation}
  G^S = -2.5\log(f^{\rm S})+C^S~,
  \label{eqZeroPoint4}
\end{equation}
{\noindent}which is related to the passband and stellar disc integrated flux $F^{S}$ in erg s$^{-1}$ as 
\begin{equation}
  f^{\rm S} = \varpi^2 ~ F^{S}~.
  \label{eqZeroPoint3}
\end{equation}
{\noindent}Our fitting technique relies on the non-linear least-squares method of Levenberg-Marquardt \citep{pressetal92-1}. The value of $\chi^2$ is taken as the sum over all passbands of the difference between both sides of Eq.\,(\ref{eqZeroPoint3}), weighted by the corresponding {\it Gaia} flux and parallax uncertainties\footnote{The error budget does not include the poorly constrained uncertainty on reddening, which has a contribution from the parallax (Eqs.\,17-19). One must be cautious when using the parameters of objects with large line of sight reddening.}. Only $T_{\rm eff}$ and $\log g$ are free parameters as the stellar radius $R$ in Eq.\,(\ref{eqZeroPoint2}) is fixed by our adopted theoretical mass-radius relation. We have performed fits with both pure-H and pure-He atmospheres for all \textit{Gaia} sources and the uncertainties on both atmospheric parameters are obtained directly from the covariance matrix.  

\subsection{The precision of {\it Gaia} atmospheric parameters}
\label{sect_atm_par.1}

For the 20\,pc sample of white dwarfs, {\it Gaia} DR2 photometry and astrometry have been used to derive effective temperatures and surface gravities \citep{hollands18}. One advantage of this volume-complete sample is that reddening is negligible. The local sample includes a wide range of spectral types, with 39 per cent or more of the remnants having a magnetic field, carbon, or metals. Despite this, \citet{hollands18} used either the pure-H or pure-He atmosphere approximation and found the {\it Gaia} photometric parameters to agree with previous photometric and spectroscopic analyses with no significant systematic offset. They found standard deviations of 3.1 per cent in \Teff\ and 0.10\,dex in $\log g$ with respect to previously published parameters. The precision of {\it Gaia} atmospheric parameters is likely better than these values owing to the inhomogeneity and lower precision of previously available ground-based observations. 

The degenerate stars in the local volume-complete sample have an average temperature of $\approx$ 8000\,K, significantly cooler than in our {\it Gaia} magnitude-limited catalogue. As a consequence, it is also important to verify the precision of {\it Gaia} atmospheric parameters for white dwarfs with $T_{\rm eff} > 10\,000$\,K, especially since the {\it Gaia} colours become increasingly less sensitive to $T_{\rm eff}$ as the latter increases \citep{carrascoetal14-1}. We have therefore employed our {\it Gaia}-SDSS spectroscopic sample (Section\,\ref{spec_SDSS_sect}) of DA white dwarfs as a reference, restricting the comparison to objects without a subtype (non-magnetic, no red excess from a companion) with a spectroscopic signal-to-noise ratio larger than 20, the latter to ensure that we do not have undetected subtypes. In addition to the available SDSS $ugriz$ photometry, we have also cross-matched this sample with the Pan-STARRS catalogue \citep{PanSTARRS}. We additionally use the bright DA stars without a subtype from \citet{gianninasetal11-1} that we have cross-matched with both our {\it Gaia} DR2 catalogue and Pan-STARRS.

For the comparison of these photometric data sets we have applied additional {\it Gaia} quality cuts (\textsc{astrometric\_excess\_noise} $<1.0$ and \textsc{phot\_bp\_rp\_excess\_factor} $<$ 1.3 + $0.06\times(G_{\rm BP}-G_{\rm RP})^2$), similar to those employed in \citet{gaiaDR2-ArXiV-5}\footnote{The comparison is also restricted to objects with best fits within the model grid and with $\sigma_{\rm Teff}/T_{\rm eff} < 0.75$ and $\sigma_{\log g} < $ 2.0.}. The Pan-STARRS $grizy$ and SDSS $ugriz$ photometry, both in the AB magnitude system, was fitted along with {\it Gaia} parallaxes using the photometric method described in Section\,\ref{sect_atm_par}. The models were integrated over the Pan-STARRS \citep{PanSTARRS2} and SDSS passbands \citep{fukugitaetal96-1}. We have applied the same reddening law as a function of wavelength and distance to all photometric data sets. At first order, reddening effects and model atmosphere systematics should not be a concern as we perform a differential comparison of multiple data sets for the same objects using the same models and reddening law. However, the {\it Gaia} passbands are considerably broader than those of SDSS and Pan-STARRS, and therefore we can not rule out these residual model effects.

\begin{figure}\includegraphics[width=0.55\columnwidth, bb = 20 175 330 570]{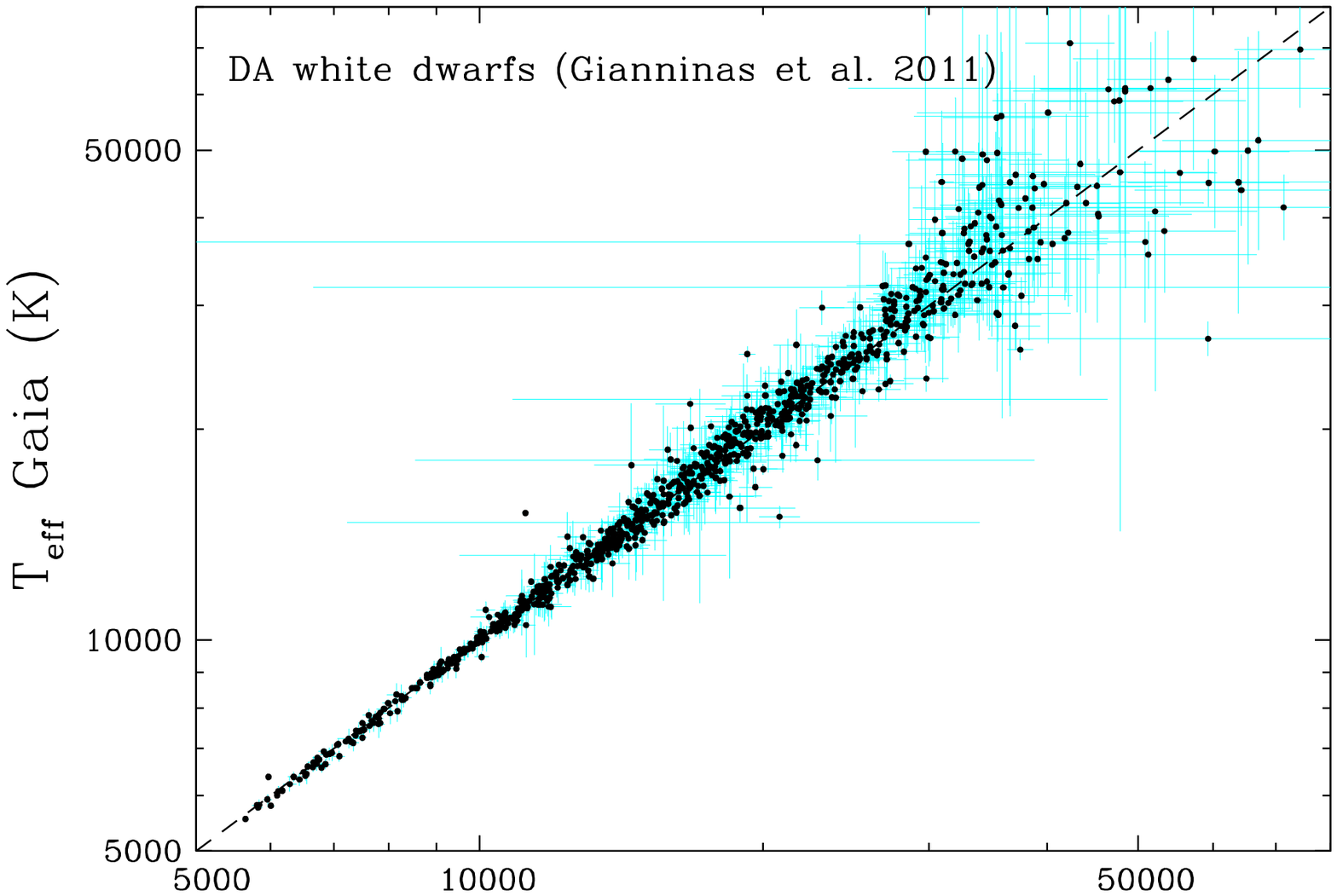}\\
\includegraphics[width=0.55\columnwidth, bb = 20 140 330 570]{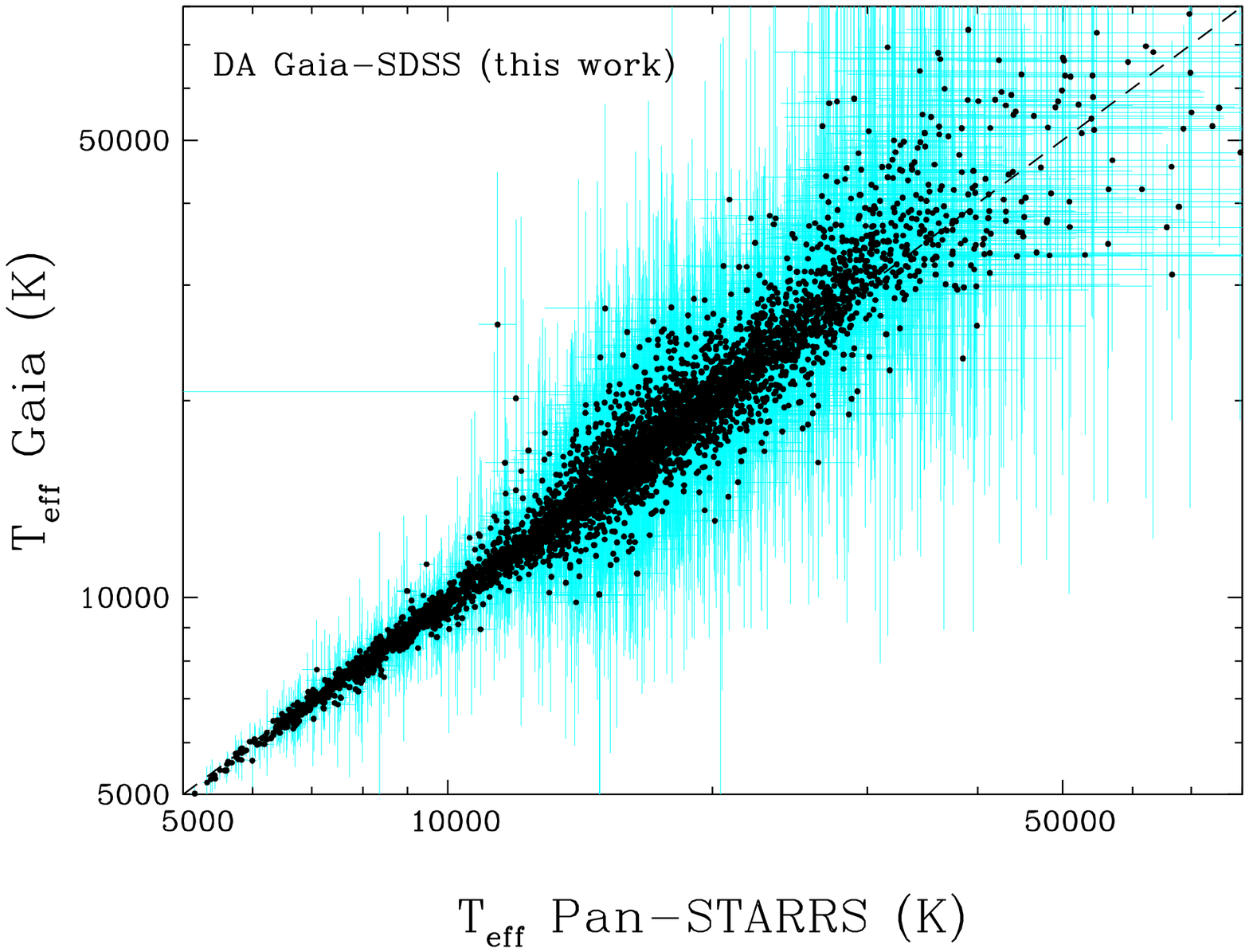}
\caption{\label{Pier_Suggestion_2} Comparison of $T_{\rm eff}$ values from photometric fits of Pan-STARRS $grizy$ and {\it Gaia} photometry for two samples of DA white dwarfs. For both photometric data sets we have employed the {\it Gaia} parallaxes. The top panel includes a sample of 1128 bright DA white dwarfs from \citet{gianninasetal11-1}, among which six objects had spurious Pan-STARRS photometry that we have replaced with data from either APASS or SDSS. The bottom panel includes 4752 DA stars from the SDSS.}
\end{figure}

\begin{figure}\includegraphics[width=0.55\columnwidth, bb = 20 175 330 570]{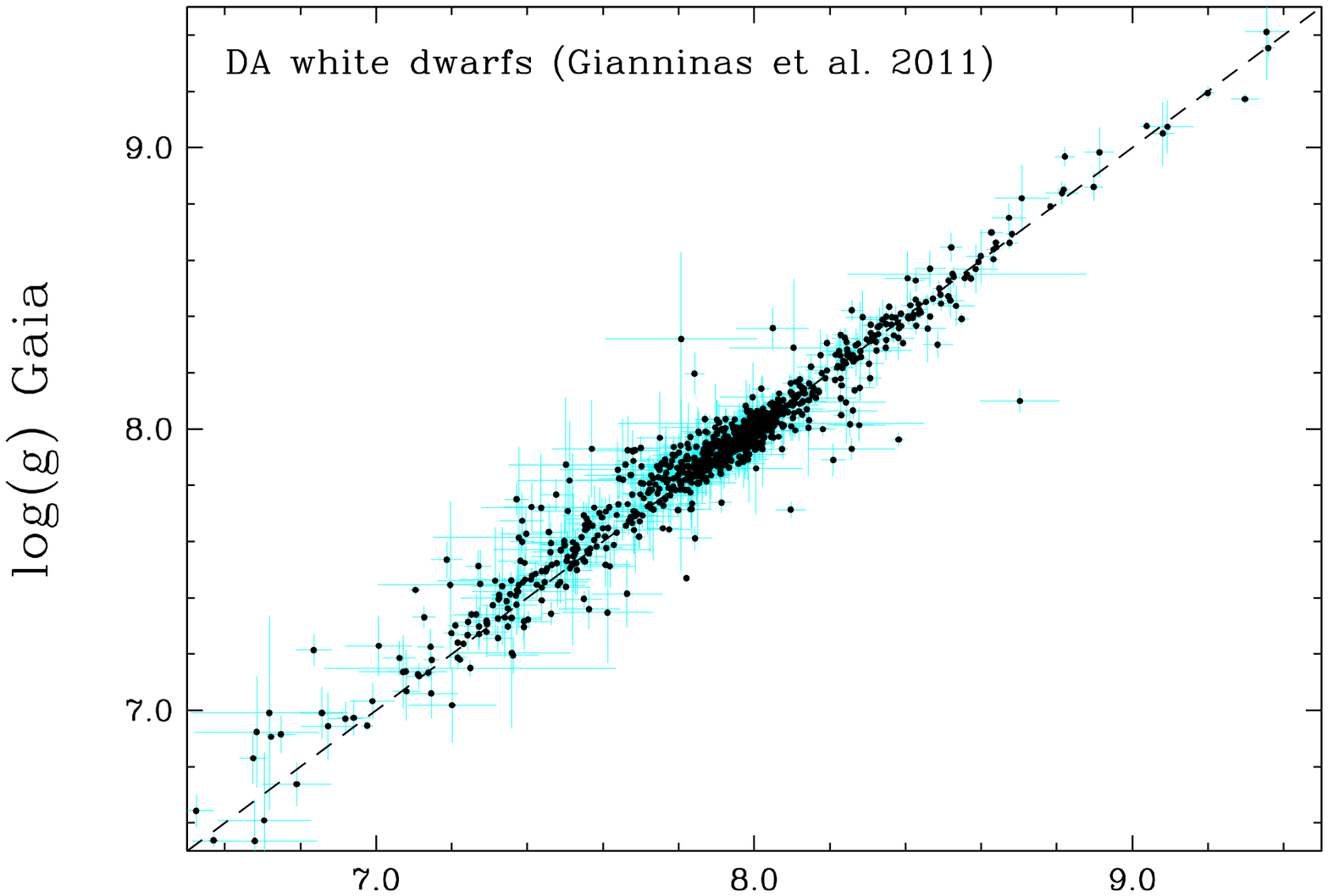}\\
\includegraphics[width=0.55\columnwidth, bb = 20 140 330 570]{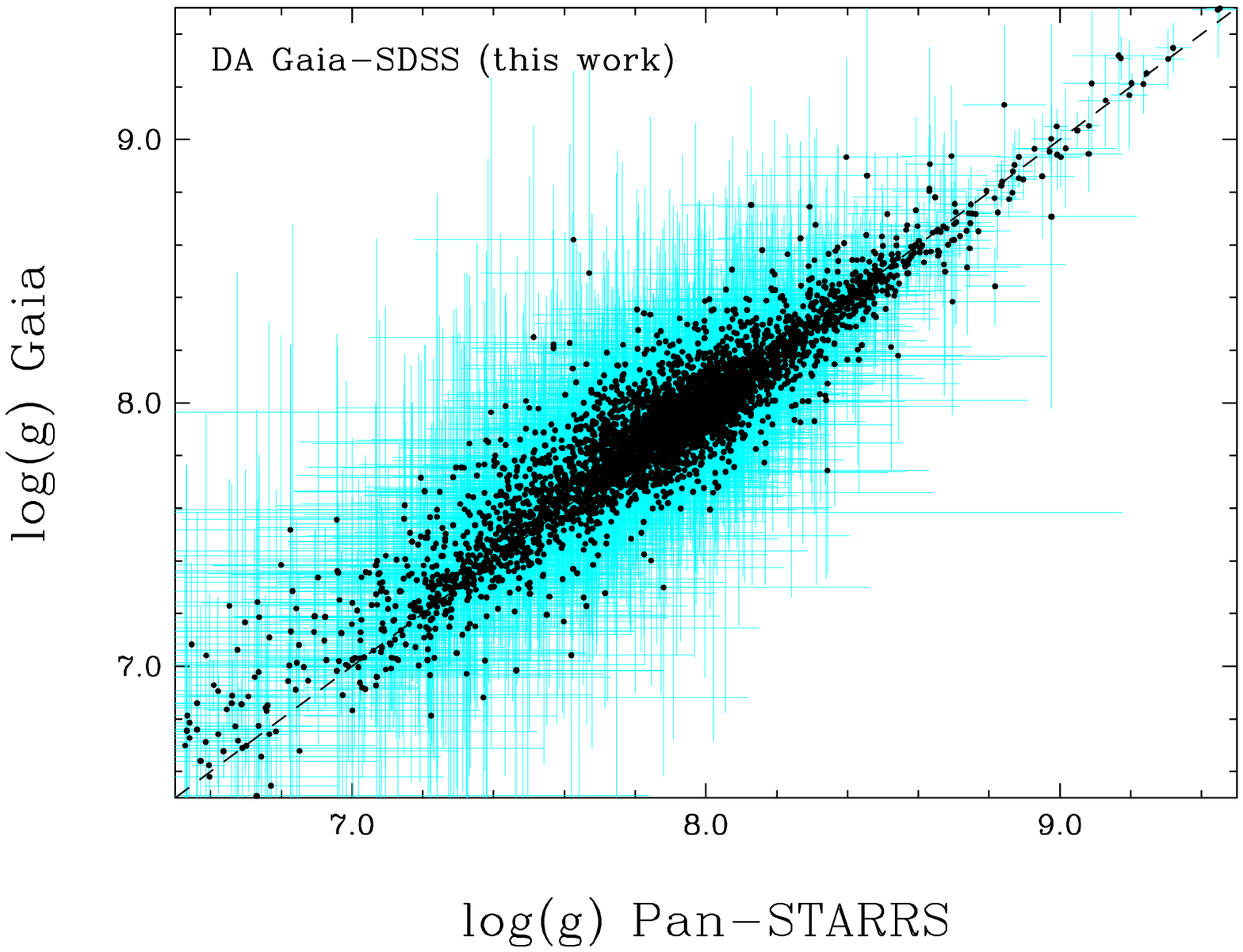}
\caption{\label{Pier_Suggestion_4} Similar to Fig.~\ref{Pier_Suggestion_2} but for the comparison of $\log g$ values from photometric fits of Pan-STARRS $grizy$ and {\it Gaia} photometry, both using {\it Gaia} parallaxes.}
\end{figure}

\begin{figure}\includegraphics[width=0.55\columnwidth, bb = 20 140 330 570]{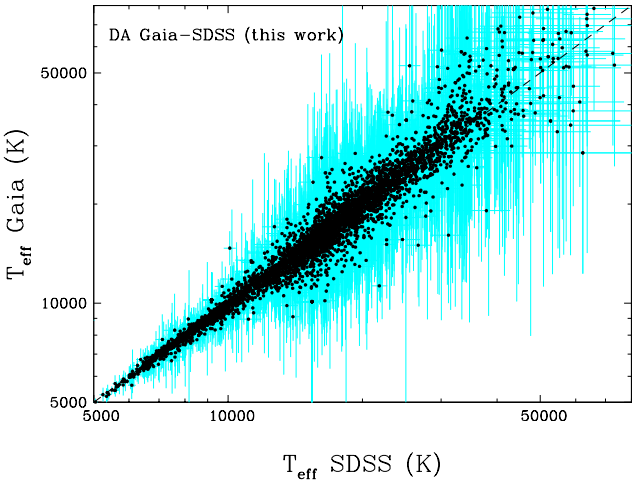}
\caption{\label{Pier_Suggestion_3} Comparison of $T_{\rm eff}$ values from photometric fits of SDSS $ugriz$ and {\it Gaia} photometry for a SDSS sample of 4778 DA white dwarfs. For both photometric data sets we have employed the {\it Gaia} parallaxes.}
\end{figure}

\begin{figure}\includegraphics[width=0.55\columnwidth, bb = 20 140 330 570]{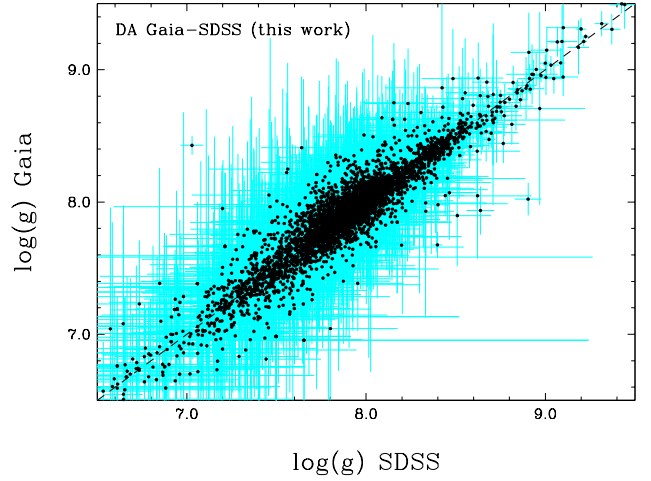}
\caption{\label{Pier_Suggestion_5} Similar to Fig.~\ref{Pier_Suggestion_3} but for the comparison of $\log g$ values from photometric fits of SDSS $ugriz$ and {\it Gaia} photometry, both using {\it Gaia} parallaxes.}
\end{figure}

The comparison of {\it Gaia} and Pan-STARRS temperatures is presented in Fig.\,\ref{Pier_Suggestion_2}. A large fraction of the brightest known DA stars are recovered in both surveys, resulting in 1128 objects from \citet{gianninasetal11-1} compared in the top panel of Fig.\,\ref{Pier_Suggestion_2}. We also show the comparison for 4778 objects which are among the brightest DA white dwarfs in the SDSS. The $\log g$ comparison is shown for both samples in Fig.\,\ref{Pier_Suggestion_4} but since we use {\it Gaia} parallaxes in all photometric fits, the shifts in surface gravities directly correspond to those in $T_{\rm eff}$ and do not provide independent information. Our results suggest that the data sets are in good agreement across the full range of $T_{\rm eff}$. Small differences in Figs.\,\ref{Pier_Suggestion_2}-\ref{Pier_Suggestion_4} could be caused by residual effects from model atmospheres or reddening given the different passbands and therefore we conclude that the {\it Gaia} photometric calibration is accurate within the combined Pan-STARRS and {\it Gaia} uncertainties. The comparison of {\it Gaia} and SDSS photometric temperatures in Fig.\,\ref{Pier_Suggestion_3} and surface gravities in Fig.\,\ref{Pier_Suggestion_5} show a similar agreement. The overlap of the \citet{gianninasetal11-1} sample with the SDSS catalogue is very small and therefore the comparison is omitted. Overall there is no evidence of any systematic offset in the {\it Gaia} photometric calibration over the 14 $\lesssim G \lesssim$ 19 magnitude range covered by our comparison samples. Our results also suggest that the SDSS and Pan-STARRS photometric data sets agree well on average with the predictions of their respective nominal passbands without the need of any empirical correction. 

\subsection{Limitations}
\label{sect_atm_par.2}

The best fit \textit{Gaia} $T_{\rm eff}$ and $\log g$ values, corresponding uncertainties and $\chi^2$ values, and implied masses from our adopted theoretical mass-radius relations, are given in columns 43-56 of our catalogue (Table\,\ref{Table format}) for both pure-H and pure-He atmospheres. The parameters are only given for a subset of the full catalogue where all of the following conditions apply 
\begin{align}
P_{\rm WD} > 0.75 ~ \textsc{or} ~P_{\rm WD}\_\textsc{flag} =1\\
\textsc{and~}\sigma_{\rm Teff}/T_{\rm eff} < 0.75\label{piercut1}\\
\textsc{and~}\sigma_{\log g} < 2.0\\
\textsc{and~}\sigma_{\rm MWD/\mathrm{M}_{\odot}} < 1.0\\
\textsc{and~}0.1 < M_{\rm WD}/\mathrm{M}_{\odot} < 1.4\\
\textsc{and~}1500 < T_{\rm eff}~{\rm[K]} < 140\,000\label{piercut2}\\
\textsc{and~}\textsc{astrometric\_excess\_noise} < 2.0\label{piercut3}\\
\textsc{and~}\textsc{phot\_bp\_rp\_excess\_factor}/\nonumber\\
  (1.3+0.06\times(G_{\rm BP}-G_{\rm RP})^2) < 1.2\\
\textsc{and~}\textsc{astrometric\_sigma5d\_max} < 2.0\label{piercut4}
\end{align}
{\noindent}with the following additional restriction for pure-He atmospheres
\begin{align}
3000 < T_{\rm eff}~{\rm[K]} < 40\,000~.
\end{align}
The restriction to high probability white dwarf candidates ($P_{\rm WD}>0.75$) is the most significant. The additional cuts remove a further 14.1 per cent of the high probability white dwarfs (see Table\,\ref{summary}) with unreliable atmospheric parameters, resulting in a final subset of 225\,370 degenerate candidates with at least one set of atmospheric parameters. The first category of cuts (Eqs.\,\ref{piercut1}-\ref{piercut2}) reflects the limited {\it Gaia} precision at fainter magnitudes and larger distances, which results in a bias predominantly against the hotter white dwarfs in the sample, but can also remove some of the most peculiar stars, such as ultra-cool white dwarfs. The second category of cuts (Eqs.\,\ref{piercut3}-\ref{piercut4}) removes strong outliers in {\it Gaia} quality flags for which the atmospheric parameters are clearly offset from the remaining objects in the sample, and therefore we have no reason to believe they are reliable. We note that it may be justified or even necessary to use further cuts on {\it Gaia} quality flags or reddening when employing the atmospheric parameters for subsamples of our catalogue.

We do not yet have spectral types for the vast majority of the {\it Gaia} white dwarfs. For the overall sample, our atmospheric parameters derived from pure-H and pure-He atmospheres agree 95.6~per cent of the time within 1$\sigma$, which suggests that the pure-H approximation may be sufficient for many applications. The $\chi^2$ values given in the table should be used with caution and we have no evidence that they can help to discriminate between spectral types. The distribution of $\chi^2$ values is fairly smooth and the tail containing large values could include both peculiar white dwarfs and objects with underestimated {\it Gaia} uncertainties. Considering the extremely small error bars of some {\it Gaia} measurements, a large $\chi^2$ value does not imply an obvious discrepancy with the input model atmospheres. 

We find an average surface gravity of $\log g$ = 8.00 assuming pure-H atmospheres. We remind the reader that while most objects with atmospheric parameters are in agreement with single star evolution, the catalogue contains a large number of sources with inferred masses below $0.46\,\mathrm{M}_{\odot}$, which would imply a main-sequence lifetime larger than the Hubble time. For the vast majority of them, the mass error is too large to rule out single star evolution. 

\section{The \textit{Gaia}-SDSS spectroscopy sample}
\label{spec_SDSS_sect}

\begin{table*}
\centering
\caption{\label{SDSS_Table} Format of the catalogue of Gaia DR2 SDSS spectra. The full catalogue can be accessed online via the VizieR catalogue access tool.}
\begin{tabular}{lll}
\hline
\hline
Column No. & Heading & Description\\
  	\hline
1 & \textsc{white\_dwarf\_name} & WDJ + J2000 ra (hh mm ss.ss) + dec (dd mm ss.s)\\
2 & \textsc{source\_id} & Unique Gaia source identifier (unique within a particular Data Release)\\
2 & \textsc{SDSS\_ra} & Right ascension of the spectrum source from SDSS DR14 [deg]\\
3 & \textsc{SDSS\_dec} & Declination of the spectrum source from SDSS DR14 [deg]\\
4 & \textsc{$u$mag} & SDSS $u$ band magnitude [mag]\\
5 & \textsc{$u$mag\_err} & SDSS $u$ band magnitude uncertainty [mag]\\
6 & \textsc{$g$mag} & SDSS $g$ band magnitude [mag]\\
7 & \textsc{$g$mag\_err} & SDSS $g$ band magnitude uncertainty [mag]\\
8 & \textsc{$r$mag} & SDSS $r$ band magnitude [mag]\\
9 & \textsc{$r$mag\_err} & SDSS $r$ band magnitude uncertainty [mag]\\
10 & \textsc{$i$mag} & SDSS $i$ band magnitude [mag]\\
11 & \textsc{$i$mag\_err} & SDSS $i$ band magnitude uncertainty [mag]\\
12 & \textsc{$z$mag} & SDSS $z$ band magnitude [mag]\\
13 & \textsc{$z$mag\_err} & SDSS $z$ band magnitude uncertainty\\
14 & \textsc{MJD} & Modified Julian date of the observation of the spectrum \\
15 & \textsc{Plate} & Identifier of the plate used in the observation of the spectrum\\
16 & \textsc{Fiber\_ID} & Identifier of the fiber used in the observation of the spectrum\\
17 & \textsc{S/N} & Signal-to-noise ratio of the spectrum calculated in the range 4500-5500 $\mathrm{\AA}$\\
18 & \textsc{spectral\_class} & Classification of the object based on a visual inspection of the SDSS spectrum\\
\hline
\end{tabular}
\end{table*}

\begin{figure*}
\includegraphics[width=1.8\columnwidth]{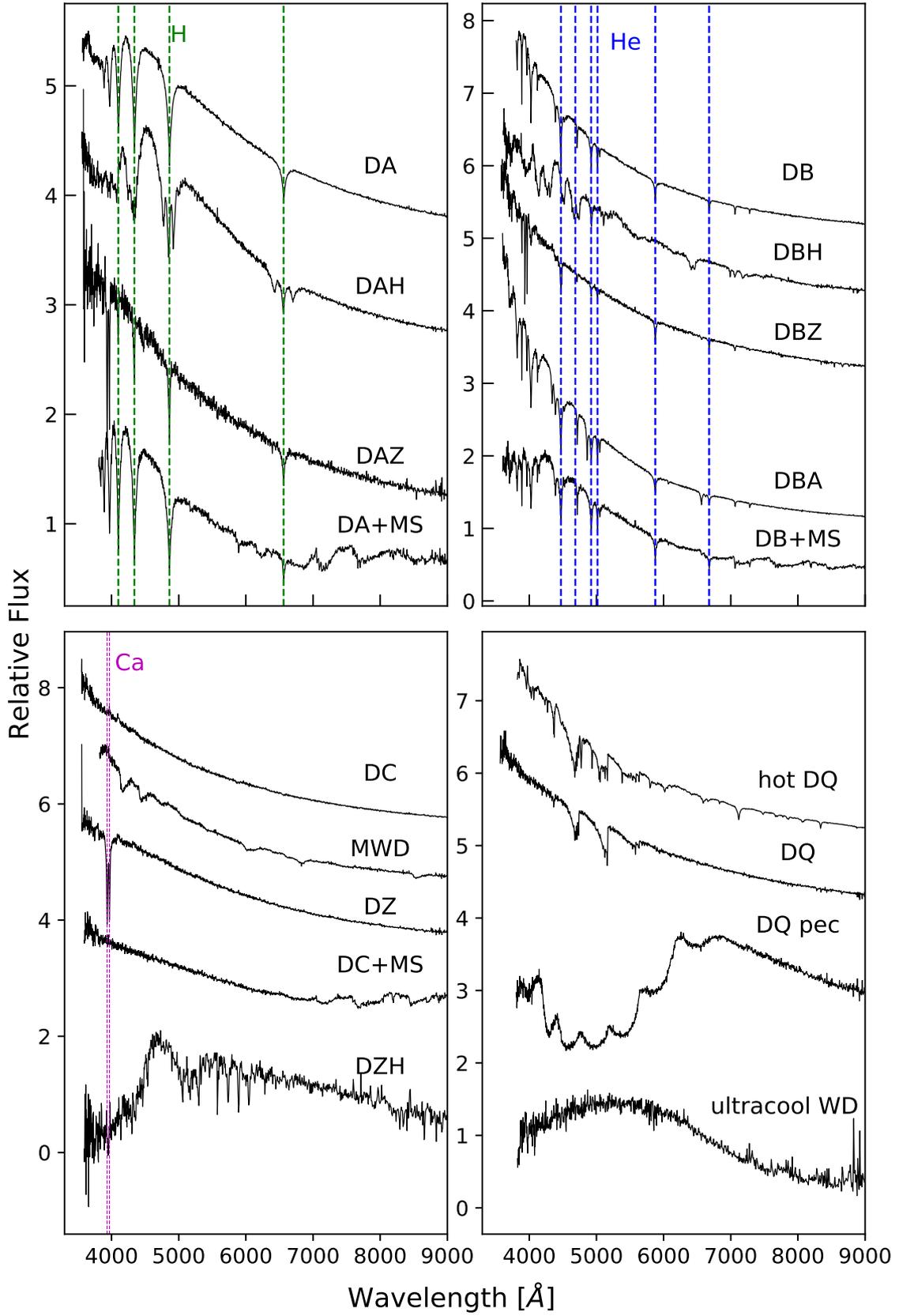}
\caption{\label{multi_spec} Representative SDSS spectra of the different white dwarf subclasses. The spectra have been offset vertically for visualisation. }
\end{figure*}

Although limited in sample size compared to the full \textit{Gaia} catalogue, the 21\,870 white dwarfs with SDSS spectroscopy currently represent the largest sample of spectroscopically confirmed \textit{Gaia} white dwarfs, and, combined with the \textit{Gaia} data, allow us to explore their global properties as well as to further characterise our catalogue.
In our classification of SDSS spectra mentioned in Section\,\ref{selection} we adopted 26 spectral types for single white dwarfs (DA, DB, DBA, DAB, DO, DAO, DC, DAZ, DZA, DBZ, DZB, DBAZ, DABZ, DZBA, DZAB, DZ, DQ, hotDQ, DQpec, DAH, DBH, DZH, MWD, PG1159, WD and WDpec, Fig.\,\ref{multi_spec}, see \citealt{sionetal83-1, koester13-1} for the definition of these  classes) and six additional classes for white dwarfs in binaries, and contaminants (CV, DB+MS, DA+MS, DC+MS, STAR, QSO).  A number of spectra have been marked as``UNKN" if we could not group them under one of the aforementioned classes, or ``Unreliable" if the quality of the spectrum was deemed to poor for visual classification. 
 Objects classified as ``MWD" are magnetic white dwarfs where line splitting is so large that we were unable to identify the atmospheric composition. While spectra marked as ``Unreliable" have a signal-to-noise ratio too low to attempt any visual classification, objects simply classified as ``WD" have spectra too poor for detailed classification, but still recognizable as those of degenerate stars. We also include a new classification for some peculiar white dwarfs as ``WDpec" (see later in this section). 

\begin{figure*}
\includegraphics[width=1.75\columnwidth]{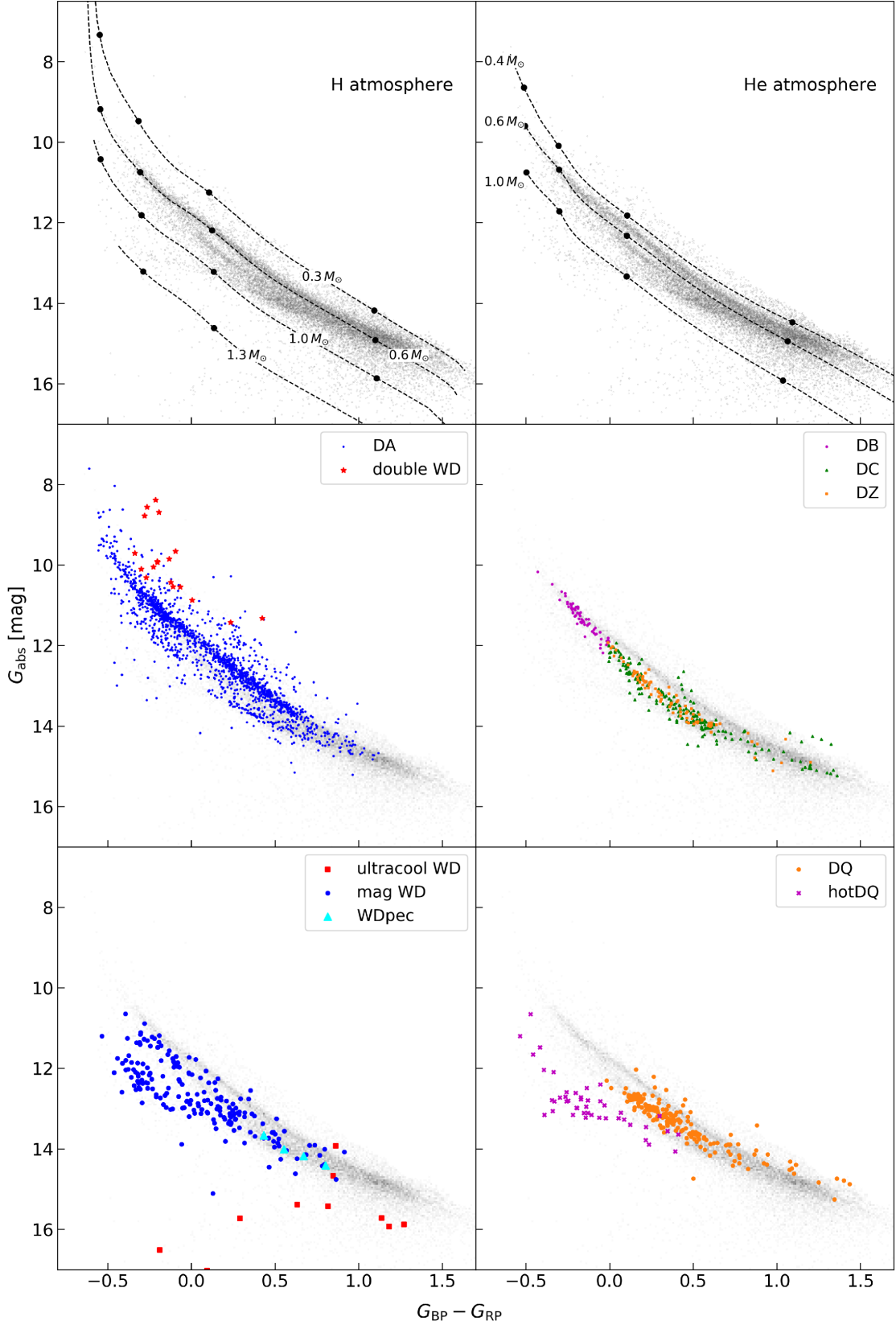}
\caption{\label{sdss_spec} \textit{Gaia} H-R diagrams showing the distribution of representative samples of various subclasses of white dwarfs. All objects were classified based on their SDSS spectra. In all panels the gray points represent the 16\,581 high-confidence white dwarf candidates from our catalogue ($P_\mathrm{wd}>0.75$ or \Pwd\_\textsc{flag}=1) within 100\,pc. 
Cooling tracks for H and He atmosphere white dwarfs at different masses are shown on the top left and top right panels,  respectively (see Section\,\ref{sect_atm_par} for a description of the evolutionary models). The black points on the cooling tracks indicate, from left to right, \Teff\ values of 40\,000\,K, 20\,000\,K, 10\,000\,K, and 5000\,K.}
\end{figure*}

In Fig.\,\ref{sdss_spec}, we display the locus of the individual white dwarf sub-classes in the \textit{Gaia} H-R diagram separately, along with the general distribution of 16\,581 white dwarf candidates within 100\,pc selected from our catalogue, adopting $P_\mathrm{wd}>0.75$ or \Pwd\_\textsc{flag}=1 (see also Section\,\ref{volume} for a discussion of the local sample). Several noticeable structures are present in the white dwarf cooling sequence, some of which have been discussed already in \citet{gaiaDR2-ArXiV-5}. The dominant feature is a bifurcation into two sequences, which are easily distinguishable at 0.0 $<\bp-\rp<$ 0.5. As illustrated in the top-left and middle-left panels, the upper one of these two tracks is easily explained as the cooling sequence of the most common DA white dwarfs ($M_{\rm WD} \approx 0.6\,\mathrm{M}_{\odot}$, see Section\,\ref{sect_atm_par} for details on the adopted evolutionary models\footnote{As discussed in \citet{gaiaDR2-ArXiV-5} and \citet{hollands18}, the observed cooling track diverges from the evolutionary models towards low masses for $T_{\rm eff} < 5000$\,K.}). In contrast, the middle-right panel in Fig.\,\ref{sdss_spec} shows that the majority of He-atmosphere white dwarfs (DB, DC\footnote{Strictly speaking, a DC classification only implies a featureless spectrum. In most cases, this is consistent with a cool He-dominated atmosphere, however, a small number of the objects classified as DC white dwarfs could have strongly magnetic H-atmospheres, wiping out the Balmer lines. }, DZ) are located on the second narrow track just below the DA $0.6\,\mathrm{M}_{\odot}$ cooling sequence. This lower branch of the bifurcation has been interpreted by \citet{kilic18} as the signature of a sub-population of high-mass white dwarfs. Even though a relatively small number of DAs also occupy this space (see Section\,\ref{He-DA}), this second track is most likely explained as the cooling sequence of canonical mass He-atmosphere white dwarfs and not as a second higher-mass sequence of H-atmosphere white dwarfs. Our conclusion is based on the empirical evidence of the spectroscopic SDSS white dwarf sample. In contrast, \citet{jimenez-esteban18-1} argued that the bifurcation cannot be based on atmosphere composition alone based on a comparison of the \textit{Gaia} H-R diagram with their population model.
We emphasise that separating the two families of objects (hydrogen dominated and helium dominated) in a clean way for the overall 100\,pc sample is practically impossible without spectroscopic confirmation of their atmospheric composition, and one should be cautious in the astrophysical interpretation of the masses assuming pure-H or pure-He atmospheres presented in Table~\ref{Table format}.

More critically, the theoretical cooling sequences for pure-He atmospheres displayed in the top-right diverge from the \textit{Gaia} observations of He-atmosphere white dwarfs for 7000\,K $\lesssim \Teff \lesssim 11\,000$\,K (0.0 $\lesssim\bp-\rp\lesssim$ 0.5). This discrepancy therefore does not impact the DB white dwarfs which largely fall onto the 0.6 $\mathrm{M}_{\odot}$ pure-He cooling sequence, but only the cooler DC, DQ and DZ stars. \citet{el-badryetal18-1} have speculated that uncertainties from additional sources of opacity in cool white dwarfs may be the cause of the diverging observed He cooling sequence, which we also conclude is the most likely explanation. However, it does not seem to have an obvious link with the presence of metals according to Fig.\,\ref{sdss_spec}. 
An additional note concerns an apparent dearth of DA white dwarfs around $\bp-\rp\simeq0.0$ (Fig.\,\ref{sdss_spec}, middle-left panel). Matching this $\bp-\rp$ colour range with the SDSS photometry (Table\,\ref{Table format})  shows that this under-density corresponds to objects with $g-r\simeq-0.2$, which is a region in colour space in which the spectroscopic completeness of SDSS is significantly reduced compared to the rest of the colour space occupied by $\Teff\ga8000$\,K white dwarfs (see Fig.\,11 of \citealt{gentilefusilloetal15-1}). We hence conclude that this particular structure in the distribution of SDSS DA white dwarfs is an artefact of the SDSS spectroscopic target selection strategy.

Comparing the cooling sequences of DA white dwarfs (middle-left panel) and those with He-dominated atmospheres (middle-right panel), it is apparent that the DA white dwarfs have a larger spread in absolute magnitudes at any given colour. The very tight sequence of the DB and DZ stars suggests that the scatter seen in the DA sequence is not a result of the larger sample size of the DA white dwarfs. On the one hand, the low-mass tail is likely to be linked to binary evolution preferentially forming DA stars \citep{gianninasetal14-1,parsons17}. The confirmed double degenerates from \citet{breedtetal17-1} and \citet{rebassa-mansergasetal17-1} are also located above the $0.6\,M_\odot$ DA cooling sequence, which is expected because the combined fluxes of the two white dwarfs make these systems intrinsically brighter. On the other hand, the mass-dependence of the mechanisms that determine the total amount of hydrogen in the envelope of white dwarfs, and how the hydrogen convectively mixes with the underlying helium layer, could explain the high-mass DA tail and the lack of massive degenerate stars with He-atmospheres \citep{kalirai05}. The initial-to-final-mass relation can also be invoked to describe the shape of that high-mass tail \citep{tremblayetal16-1,el-badryetal18-1}. Following upon the investigation of \citet{kalirai05}, we note that 64 out of the 65 white dwarfs that are confirmed young open cluster members (cluster age $<$ 700 Myr) in \citet{cummings16} are DA stars (the one DBA star is the Hyades member WD\,0437+138). These objects cover the range $M_{\rm WD} > 0.65\,\mathrm{M}_{\odot}$ and $M_{\rm initial} > 2.5\,\mathrm{M}_{\odot}$, with 69 per cent of the sample below the so-called DB gap or deficiency \citep[$T_{\rm eff} \lesssim 30\,000$\,K;][]{bergeronetal11-1,koesteretal15-1} for field white dwarfs. This provides strong evidence that single star evolution can explain the lack of massive DB stars.

The DA sample also appears to have an over-density of under-luminous stars below the $M_{\rm WD}=0.6\,\mathrm{M}_{\odot}$ cooling track forming a separate third sequence (distinguishable at 0.0 $\lesssim\bp-\rp\lesssim$ 1.0 in the middle left panel of Fig.\,\ref{sdss_spec}), located below the cooling track of He-rich white dwarfs discussed above. This ``transversal" sequence, also seen in the overall 100\,pc sample (top panels), does not run parallel to the DA cooling sequences and is therefore not a constant mass track, ruling out a straightforward astrophysical explanation such as binary evolution or effects from the initial-to-final mass relation. Explaining the origin of this feature is beyond the scope of this paper but we speculate that it could be the result of a mass-dependent cooling effect (Tremblay et al. 2018, subm.).

\begin{figure}
\includegraphics[width=1.\columnwidth]{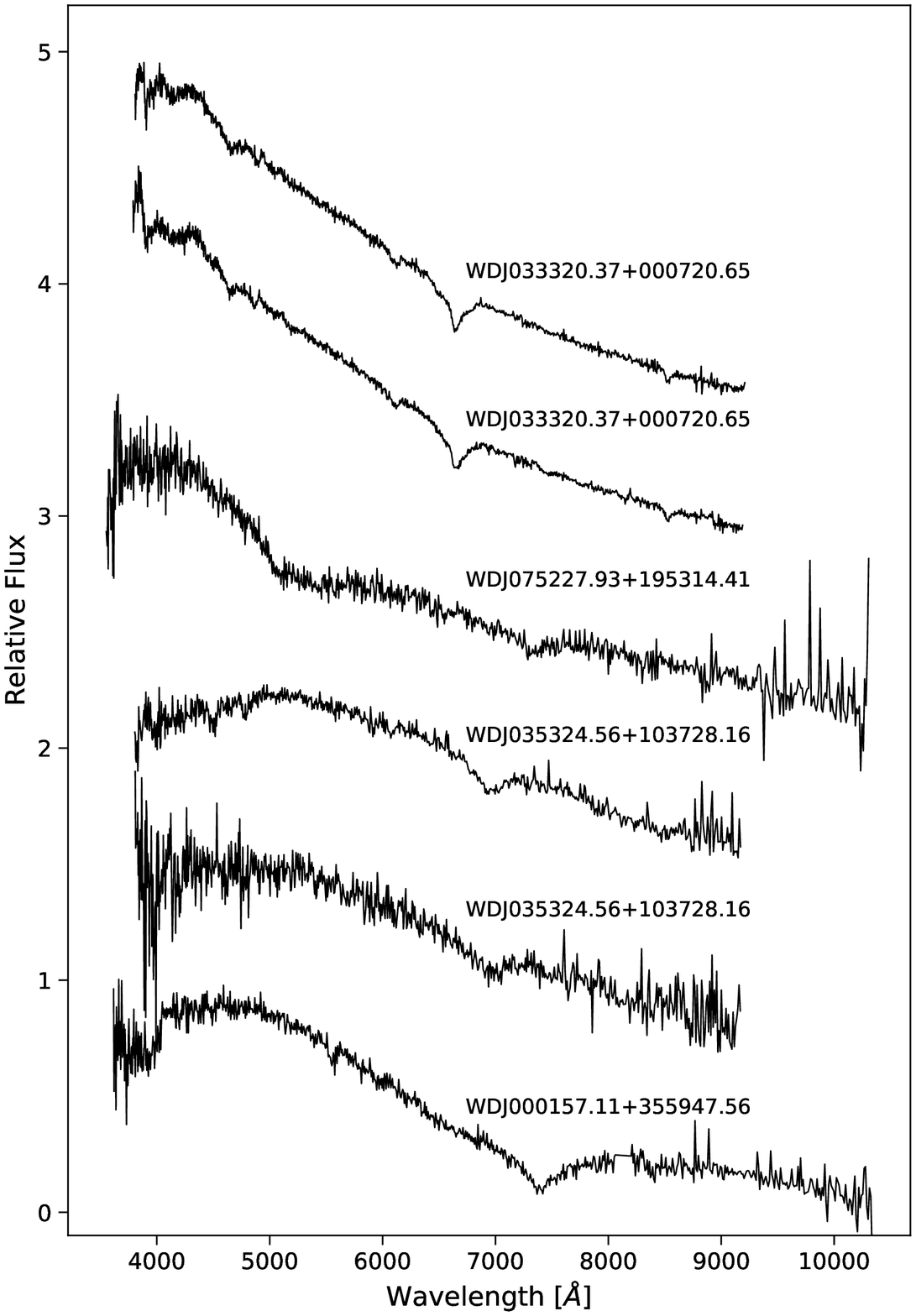}
\caption{\label{strange_spec} All available SDSS spectra of four peculiar white dwarfs (WDpec).}
\end{figure}

In the bottom left panel of Fig.\,\ref{sdss_spec} we show the location of a representative number of magnetic white dwarfs. It appears that these objects span a relatively large range of absolute magnitudes for a given colour, but on average they are under-luminous compared to typical DA white dwarfs. This finding seems to corroborate the long standing theory that these white dwarfs are more massive and so smaller than their non-magnetic counterparts \citep{liebertetal88-1, ferrarioetal15-1}. 
In our visual inspection of SDSS spectra we also identified four objects that despite having a parallax and colours consistent with those of white dwarfs, have a unique spectral appearance among the 21\,870 white dwarfs spectroscopically confirmed by SDSS (Fig.\,\ref{strange_spec}). These peculiar white dwarfs (classified as WDpec in our catalogue) all exhibit one broad absorption feature and a number of smaller ``satellite" absorption lines. The main broad absorption features appear to be shifted by hundreds of \AA\ from star to star.  In the H-R diagram these four stars line up below the DA $0.6\,\mathrm{M}_{\odot}$ cooling sequence much like most of the magnetic white dwarfs. Two of these stars (WD\,J033320.57+000720.65 and WD\,J075227.93+195314.41) are already known as magnetic degenerates with unidentifiable features \citep{reimersetal98-1, kepleretal15-1}, so we speculate that these peculiar objects may all be members of the same family of magnetic white dwarfs. 
We are however unable to venture any hypothesis on their atmospheric composition.

Fig.\,\ref{sdss_spec} (bottom-left panel) also illustrates the location of spectroscopically confirmed ultracool white dwarfs (\Teff\ $\lesssim 4000$\,K) \citep{harrisetal01-2, gatesetal04-1,harrisetal08-1, hollandsetal17-1}. These  objects, still rare even in the very large \textit{Gaia} sample, migrate to bluer colours (and so a distinct location compared to hotter white dwarfs) as a result of collision-induced absorption \citep{hansen98}. Many of these white dwarfs occupy areas of the H-R diagram in which we apply the \textsc{\Pwd\_flag} (see Section~\ref{selection}, Table\,\ref{Table format}), so particular care should be taken when attempting to select these objects from the catalogue.
Finally the bottom right panel shows a distinction in the distribution of DQ \citep{dufouretal05-1} and hot DQ white dwarfs \citep{dufouretal07-1}. Cooler DQ stars roughly line up with the cooling sequence of other He-atmosphere white dwarfs, while hot DQs appear distinctly under-luminous and occupy the same locus as many magnetic white dwarfs. This is not surprising, as a large fraction, if not all, hot DQs are thought to harbour magnetic fields \citep{dufouretal08-1, lawrieetal13-1, williamsetal13-1,williamsetal16-1}. We note that hot DQ stars, and in general the entire lower branch of the magnetic white dwarf cooling track, overlap with the ``transversal" sequence observed for DA white dwarfs.

\subsection{He-atmosphere DA white dwarfs}
\label{He-DA}
\begin{figure}
\includegraphics[width=0.95\columnwidth]{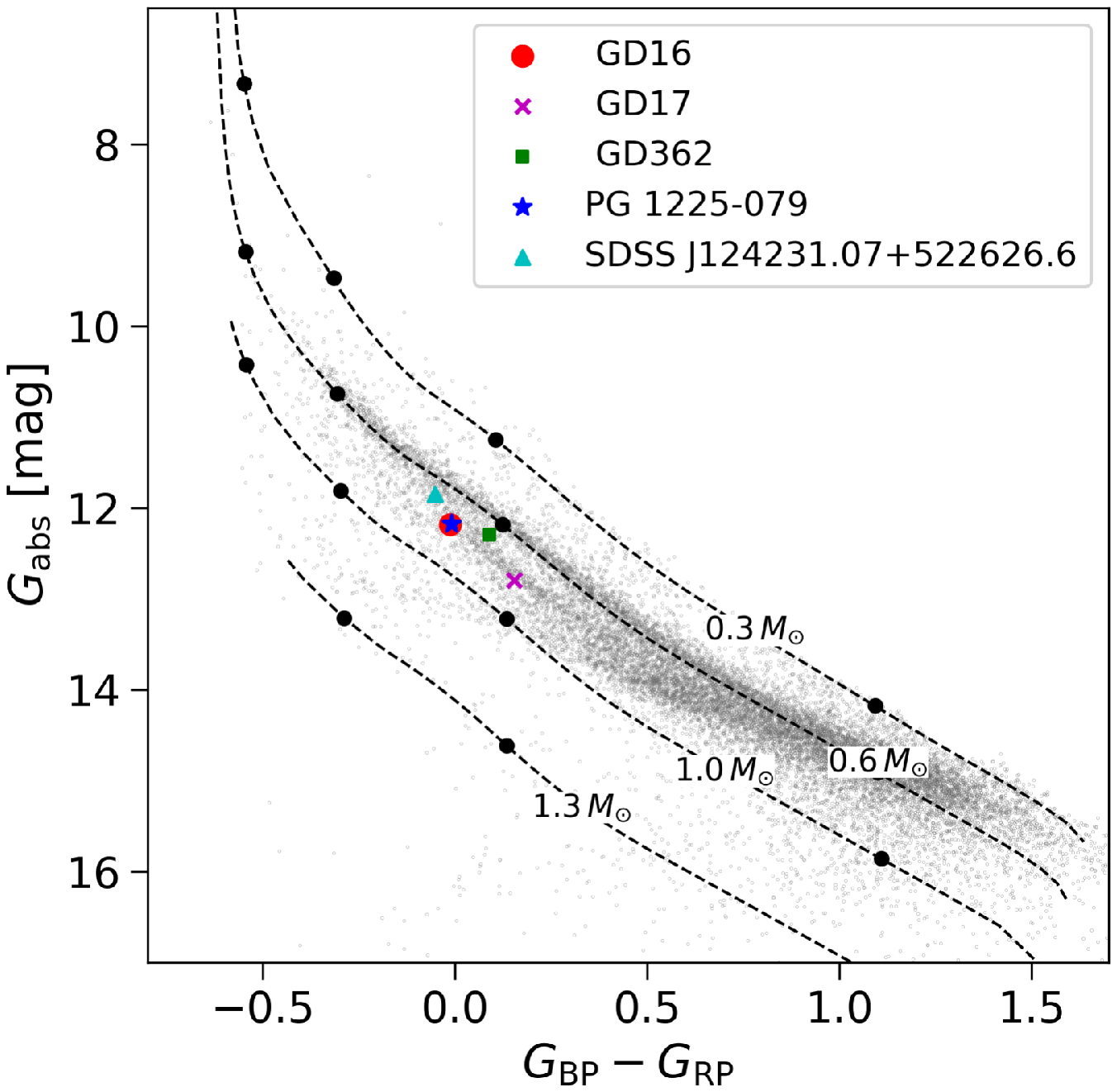}
\caption{\label{DA_He}\textit{Gaia} H-R diagram showing the distribution of five known He-atmosphere DAZ white dwarfs \citep{koesteretal05-1, gentilefusilloetal17-1, gianninasetal04-1, kawka+vennes05-1, kilkenny86-1, raddietal15-1}. GD16 and PG~1225$-$079 overlap on the panel. The gray points represent our 100\,pc sample of white dwarf candidates. Cooling tracks for H atmosphere white dwarfs at different masses are shown on the top left and top right panels, respectively (see Section\,\ref{sect_atm_par} for a description of the  evolutionary models).}
\end{figure}

A small number of stars with the spectroscopic appearance of DAZ white dwarfs are known to actually have He-dominated atmospheres with unusually large H components and metal pollution. In Fig.\,\ref{DA_He} we show the position of the currently known five members of this family: GD\,16, GD\,17, GD\,362, PG\,1225$-$079, and SDSS\,J124231.07+522626.6 \citep{koesteretal05-1, gentilefusilloetal17-1, gianninasetal04-1, kawka+vennes05-1, kilkenny86-1, raddietal15-1} on the observed \textit{Gaia} H-R diagram. All five objects broadly lie on the observed \textit{Gaia} cooling sequence of He atmosphere white dwarfs. This could indicate that these peculiar objects evolve in a similar way as average mass He-atmosphere white dwarfs.  Alternatively, these objects could behave as thin-H layer DA white dwarfs without suffering from the bifurcation problem of He-atmospheres, and therefore have masses slightly higher than the canonical $0.6\,\mathrm{M}_{\odot}$. If the location on the He-atmosphere sequence were to be confirmed for other He-rich DA white dwarfs, this property could be exploited to identify more of these objects, and help to  unravel the question of the origin of the H in their atmosphere. Indeed as shown in Fig.\,\ref{sdss_spec} a number of spectroscopically confirmed SDSS DA white dwarfs occupy the same region on the H-R diagram and analogously to these five metal polluted stars, some may actually have He dominated atmospheres, especially if their spectroscopic masses assuming pure-H atmospheres are unusually large \citep{tremblayetal10-1,rolland18}.

\section{Discussion}
\subsection{Sky density and limiting magnitude}
\label{sky_info}
\begin{figure}
\includegraphics[width=0.95\columnwidth]{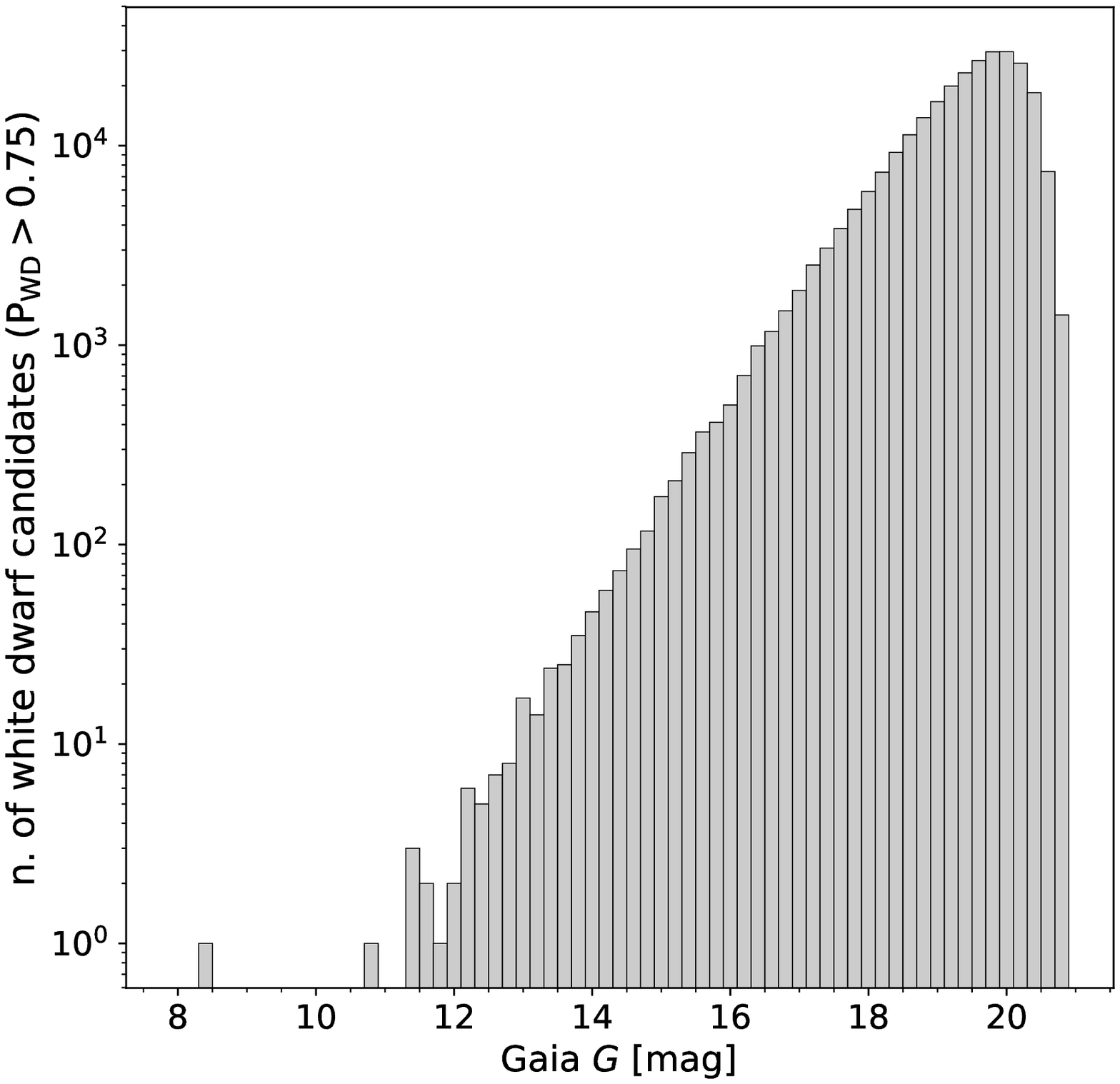}
\caption{\label{G_hist} Number of high-confidence white dwarf candidates ($\Pwd > 0.75$) as function of \textit{Gaia} $G$ magnitude.}
\end{figure}

\begin{figure}
\includegraphics[width=0.95\columnwidth]{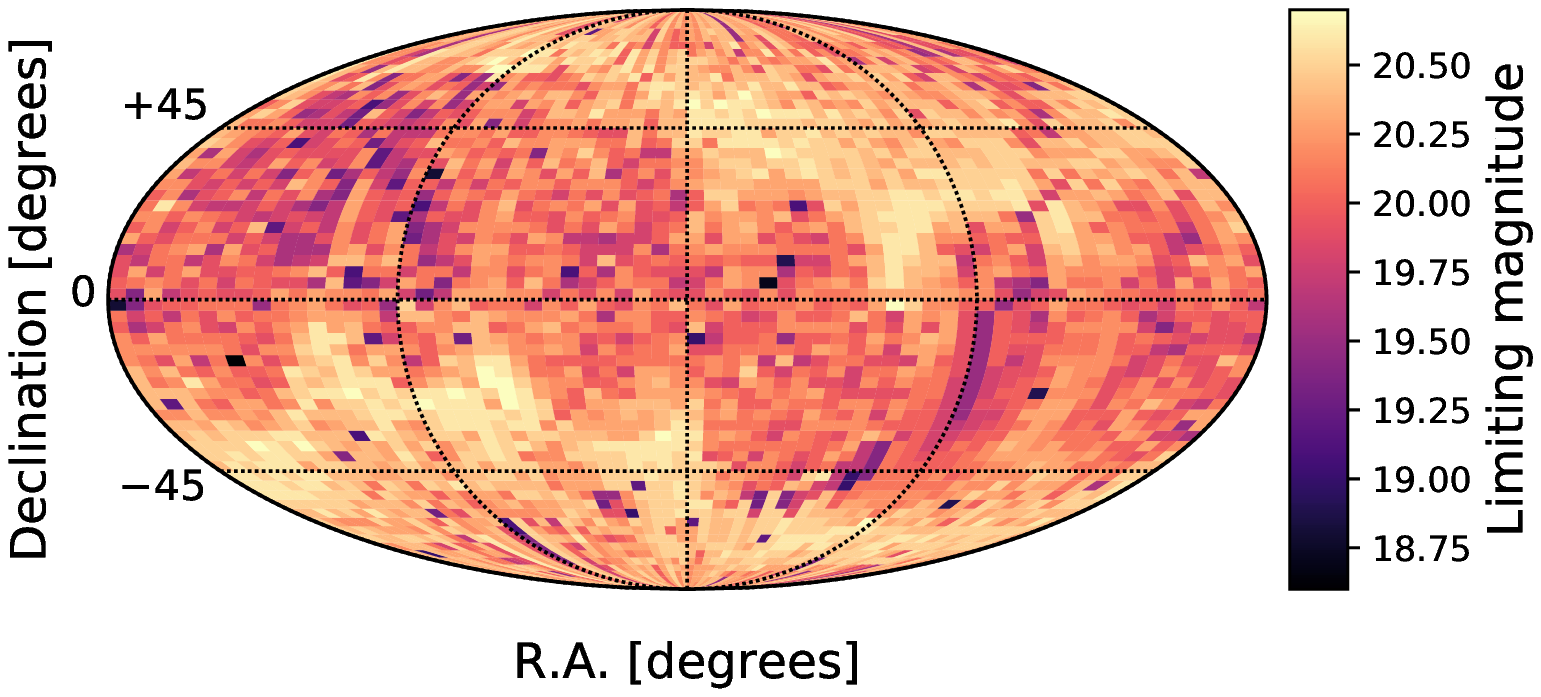}
\caption{\label{Gaia_limit} Limiting magnitude for \textit{Gaia} white dwarfs calculated using 10 deg$^{2}$ bins.}
\end{figure}

Using a \Pwd\ $>0.75$ reference sample we can attempt to estimate the overall sky density of  white dwarfs in \textit{Gaia} DR2. Contrary to what is expected from simple Galactic structure, the sky density of white dwarfs in our catalogue does not significantly increase at lower Galactic latitudes (Fig.\,\ref{Gaia_allsky}). This is a consequence of the stricter selection we apply to the areas with high stellar densities at lower Galactic latitudes (Section\,\ref{selection}). Additionally, virtually no white dwarfs are found in the most central regions of the plane where crowding is highest ($300\gtrsim l \lesssim40$, $|b|\lesssim6$). We find significant structure in the density of white dwarfs across the entire sky, as a result of the non-uniform limiting magnitude of \textit{Gaia} observations.
Though the overall limiting magnitude of our catalogue is $\simeq 20.1$ (Fig.\,\ref{G_hist}), it can vary by more than 1\,mag across the sky in a pattern that closely follows that of \textit{Gaia} scanning law (Fig.\,\ref{Gaia_limit}).  In DR2, a limiting magnitude of least 20 is reached for 75 per cent of the sky, and we estimate the sky density of white dwarfs in these regions with $G \leq 20$ to be $\simeq 4.5~\mathrm{deg}^{-2}$.
We can assume that with future \textit{Gaia} data releases the effective limiting magnitude will become more uniform across the sky, and in subsequent versions of our catalogue we will be able to identify more faint white dwarfs.

\subsection{Comparison with an SDSS sample of white dwarf candidates}
\label{pwd_compare}

\begin{figure}
\includegraphics[width=0.95\columnwidth]{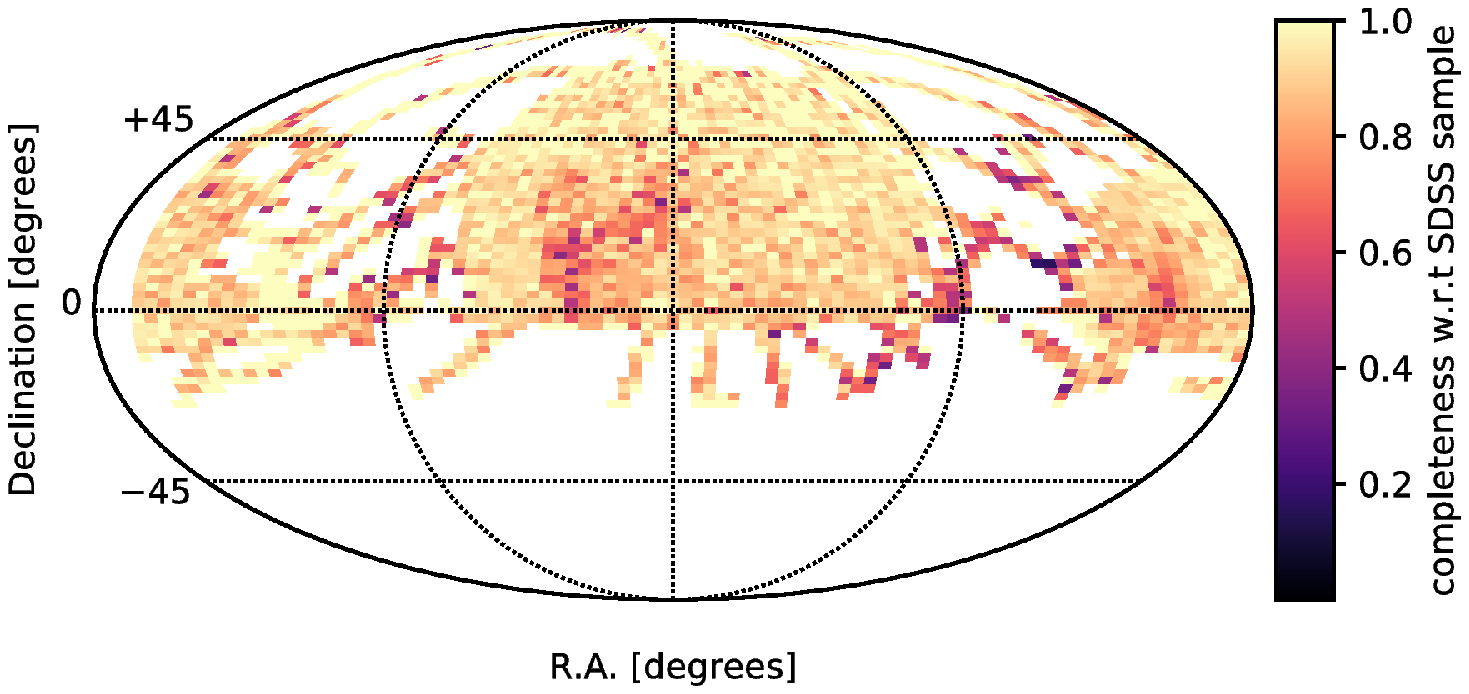}
\caption{\label{sky_hist} Completeness of our \textit{Gaia} catalogue of white dwarf candidates with respect to the SDSS comparison sample, as a function of sky position. Each bin represents 5 deg$^{2}$.}
\end{figure}

\begin{figure}
\includegraphics[width=0.95\columnwidth]{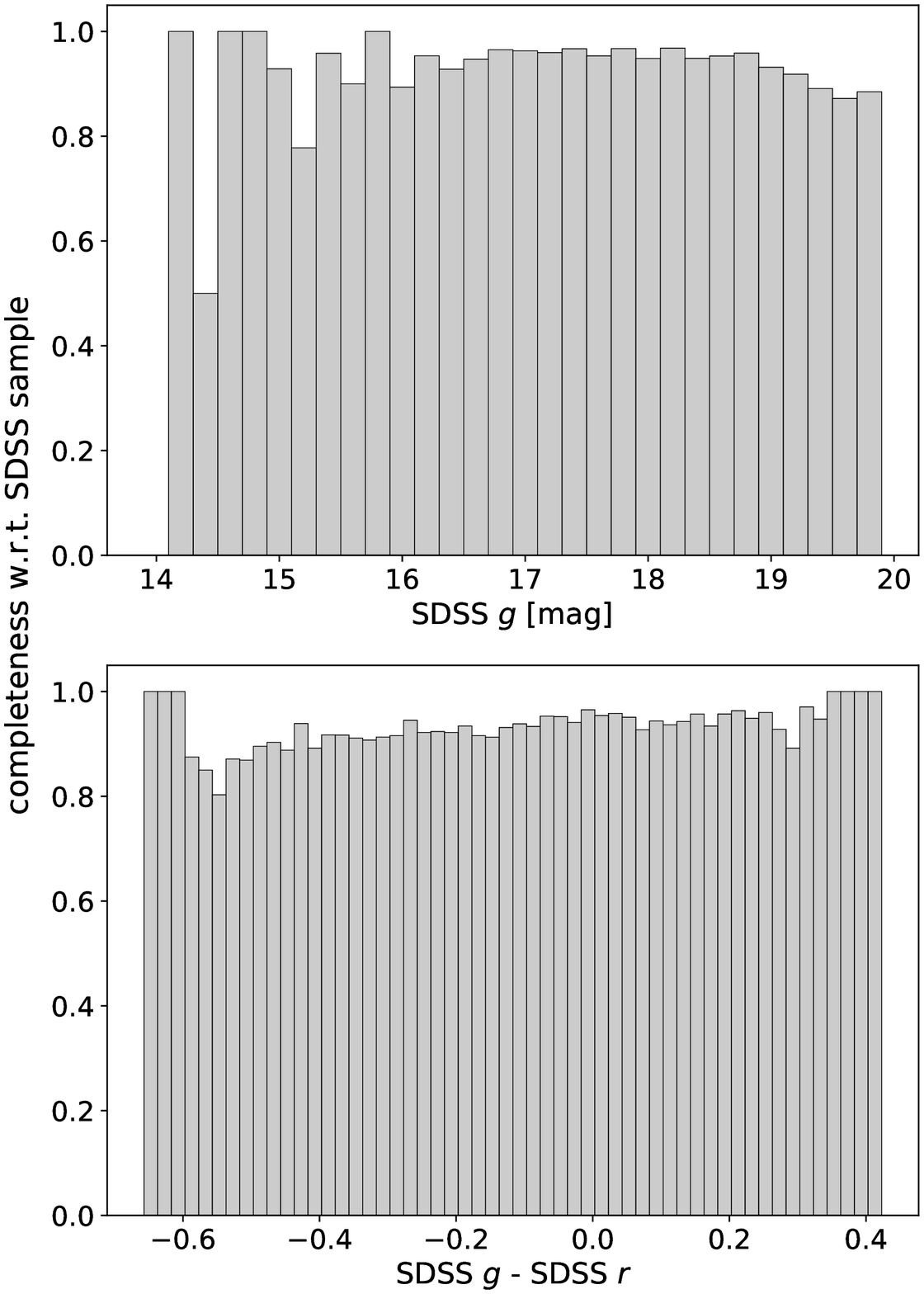}
\caption{\label{mag_hist} Similar to Fig.\,\ref{sky_hist}, but for the completeness as a function of magnitude and colour. }
\end{figure}

With $\simeq 260\,000$ high-confidence candidates, our catalogue of white dwarfs is certainly the largest ever published, but in order to explore the full diagnostic potential of this vast sample, we need to evaluate the completeness of our selection. A number of factors within \textit{Gaia}\,DR2 and/or in our  selection method may cause some genuine white dwarfs to be excluded from this catalogue. In order to assess this issue it is necessary to compare the \textit{Gaia} catalogue of white dwarfs with a sufficiently large and well characterised sample of stellar remnants. The spectroscopic samples of white dwarfs currently available (e.g., SDSS in Section\,\ref{spec_SDSS_sect}) are ill-fitted for this task as they are severely incomplete and biased by the specific observing strategy adopted. Therefore we decided to rely on a sample of SDSS white dwarf candidates selected on the basis of their colour and reduced proper motion as described in \citet{gentilefusilloetal15-1}. However, the original catalogue of \citet{gentilefusilloetal15-1} only included objects brighter than $g=19$ as fainter sources did not have reliable proper motions in SDSS. In order to create a sample which better matches the magnitude limit of our \textit{Gaia} catalogue we extended the \citet{gentilefusilloetal15-1} catalogue to $g\leq20.1$ by making use of the more accurate proper motions from the Gaia-PS1-SDSS (GPS1) catalogue \citep{tianetal17-1}. Full details about the development and characterisation of this deep SDSS comparison sample are available in Appendix A. 

From this deep photometric SDSS catalogue we select a sample of  60\,739 white dwarf candidates, which only has seven per cent contamination while still including 97 per cent of all the white dwarfs in the full sample. However, it is important to notice that, because of the colour restrictions used, this sample only contains white dwarfs with \Teff\ $>7000$\,K, and an additional $\simeq 14\,000$ stellar remnants in the SDSS footprint are potentially missing because they have no proper motion measurement.  For completeness, we note that the footprint of the SDSS photometry mostly covered high Galactic latitudes with $|b|\gtrsim20^{\circ}$.
In conclusion, we estimate the deep SDSS sample to contain $\simeq 75$ per cent of all the white dwarfs observed by SDSS, brighter than $g=20.1$ and with \Teff\ $>7000$\,K. 

We cross matched our \textit{Gaia} catalogue of white dwarf candidates with the deep SDSS comparison sample and retrieved 47\,503 of the SDSS white dwarf candidates. Accounting for the expected level of contamination of the deep SDSS sample (7 per cent), we can use the percentage of objects missing in the \textit{Gaia} white dwarf candidate sample to estimate an upper limit in completeness of the \textit{Gaia} catalogue of 85~per cent for white dwarfs with $G$ $\leq20$ and \Teff\ $> 7000$\,K, at high Galactic latitudes ($|b|>20^{\circ}$). 
Similarly, we can use the estimated completeness of the deep SDSS sample and the number of objects we retrieved in the cross-match with the {\it Gaia} catalogue to calculate a lower limit in completeness of 60 per cent. Additionally, we can use this comparison as a diagnostic of potential biases in our \textit{Gaia} selection. As illustrated in Fig.\,\ref{sky_hist} the completeness of our \textit{Gaia} catalogue drops close to the Galactic plane and in these areas the upper limit on the overall completeness can be as low as 50 per cent. This effect is a direct consequence of the stricter quality selection we impose on crowded areas at low Galactic latitudes (see Section\,\ref{selection}). Even in the era of \textit{Gaia} the Galactic plane represents a challenging environment to be surveyed accurately, nonetheless the catalogue presented here still includes the largest sample of Galactic plane white dwarf candidates available to date. A potentially more complete selection of white dwarfs in the Galactic plane could be achieved combining \textit{Gaia} observations with dedicated photometric surveys (e.g., IPHAS; \citealt{drewetal05-1} or VPHAS+; \citealt{vphas+14-1}).

We also tested the relative completeness of our \textit{Gaia} white dwarf selection as a function of magnitude and colour (Fig.\,\ref{mag_hist}). 
Since the level of contamination of our deep SDSS comparison sample is itself colour dependent, for this test we use a sample of $\simeq 13\,000$ high signal-to-noise ratio spectroscopically confirmed SDSS degenerate stars. This comparison does not reflect the absolute completeness of our \textit{Gaia} catalogue and should only be used to explore any potential correlation with magnitude and/or colour. Fig.\,\ref{mag_hist} (top panel) shows no obvious correlation with magnitude. The apparent drop in completeness at $g<15$ is most likely due to small number statistics as a consequence of SDSS reaching saturation.
 
The bottom panel in Fig.\,\ref{mag_hist} clearly illustrates that there is no marked colour trend in the completeness of our catalogue with respect to the SDSS spectroscopic sample. However, our spectroscopic comparison sample only includes white dwarfs with \Teff\ $> 7000$\,K and the completeness of our \textit{Gaia} selection may vary for cooler (and redder) objects.

We emphasize that the completeness values we estimate refer to the magnitude limited sample and so across all distances. The vast majority of the SDSS white dwarf candidates that we do not recover are faint and distant stars with relatively poor \textit{Gaia} measurements. The completeness for volume limited samples is therefore likely much higher than these values as our selection is not biased in either colour or apparent magnitude (Fig.\,\ref{mag_hist}). This statement is corroborated by the fact that all of the 20\,pc white dwarfs identified by \citet{hollands18} are included in our catalogue.

\subsection{Volume completeness}
\label{volume}

\citet{hollands18} carefully determined the selection function and completeness of the {\it Gaia} DR2 sample of 139 white dwarfs within 20\,pc and found the space density to be $(4.49\pm0.38)\times10^{-3}$\,pc$^{-3}$. To recover these numbers with our catalogue, we must apply $P_{\rm WD} > 0.75$ and $\textsc{astrometric\_excess\_noise} < 1.0$ owing to the relatively large number of main-sequence stars scattered in the local sample owing to erroneous parallaxes \citep{hollands18}. 

This can be compared with the number of degenerate stars that we find at larger distances in our catalogue, and here we take a particular interest in the 100\,pc sample. First, we must consider the quality cuts to apply. There are 16\,581 white dwarf candidates in our catalogue with $\varpi > 10$ mas and $P_{\rm WD} > 0.75$ or \Pwd\_\textsc{flag}=1, though a considerable number of those have poor quality flags. We employ the sample of white dwarfs with reliable atmospheric parameters (Section \ref{sect_atm_par.2}) as a compromise, resulting in 15\,109 high-confidence members of the 100\,pc sample, and a  cut of 8.9 per cent compared to the initial sample. The number of objects removed is similar to the cut $\textsc{astrometric\_excess\_noise} < 1.0$ (8.3 per cent). The inferred space density within 100\,pc is  86.9 per cent of that found for the 20\,pc sample, though at such large distances the approximation of constant space density is unlikely to hold because of the finite scale height of the Galactic disc. Furthermore, the white dwarf luminosity function is known to peak at $G_{\rm abs} \approx 15-16$ before dropping off at fainter magnitudes owing to the finite age of the disc \citep{winget87}. Given the sky average limiting {\it Gaia} $G$ magnitude of $\approx$ 20 for our white dwarf catalogue, it is unlikely to be complete for distances larger than about 60\,pc \citep[see also][]{carrascoetal14-1}.

\begin{figure}\includegraphics[width=0.55\columnwidth, bb = 10 120 280 600]{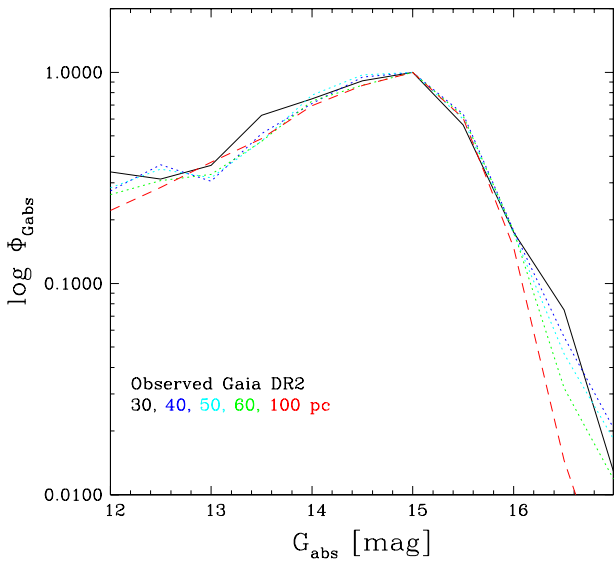}
\caption{\label{luminosityfunction} {The \it Gaia} DR2 white dwarf absolute $G$ magnitude function for limiting distances of 30, 40, 50, 60, and 100\,pc, normalised at $G_{\rm abs}$ = 15.}
\end{figure}

Fig.\,\ref{luminosityfunction} presents the normalised {\it Gaia} white dwarf absolute $G$ magnitude distribution (analogous to a luminosity function) for different limiting distances. The similarity of the functions from 30 to 50\,pc confirms that our catalogue is essentially a volume-limited sample up to that distance. At larger distances, the number of cool and/or massive degenerate stars with $16 < G_{\rm abs} < 17$ is clearly decreasing as a result of the {\it Gaia} limiting magnitude. Nevertheless, the drop in the absolute magnitude function is still fairly small and it is not expected to impact the total number of white dwarfs within 100\,pc by more than a few per cent.

\begin{figure}\includegraphics[width=0.55\columnwidth, bb = 10 120 280 600]{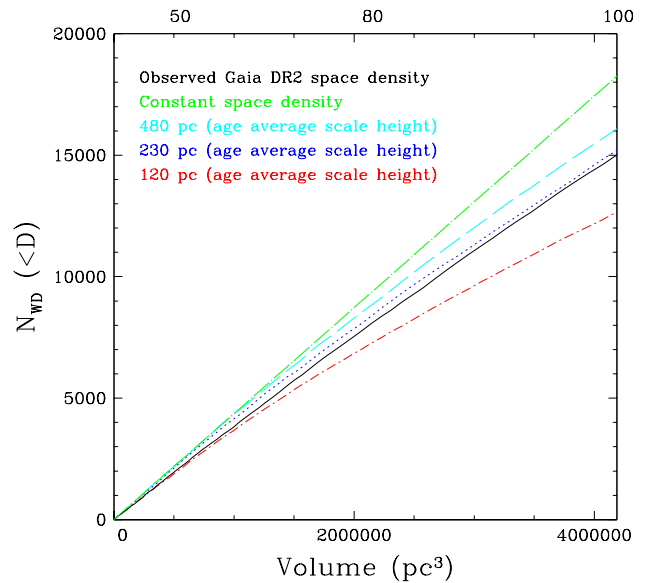}
\caption{\label{completeness3} Cumulative number of white dwarfs as a function of volume for a clean subsample of our {\it Gaia} catalogue (solid black). This is compared with the constant space density as inferred from the 20\,pc sample (green, dot long-dashed). We also show our population synthesis predictions assuming a magnitude limit of $G = 20$\,mag and an age average vertical scale height of 480\,pc (cyan, long-dashed), 230\,pc (blue, dotted), and 120\,pc (red, dot short-dashed). All simulations are normalised to the 20\,pc space density. The approximation of constant space density does not account for the fact that the faintest white dwarfs within 100\,pc can not be detected with {\it Gaia}.}
\end{figure}

To understand further the properties of the 100\,pc sample, we employ the white dwarf population simulation drawn from \citet{tremblayetal16-1}. In brief, this simulation uses constant stellar formation history over the past 10 Gyr, the Salpeter initial mass function, main-sequence lifetimes for solar metallicity from \citet{Hurley00}, the initial-to-final mass relation of \citet{cummings16}, a uniform distribution in Galactic coordinates U and  V (corresponding to the plane of the disc), and Galactic disc heating in the vertical coordinate W \citep{seabroke07} starting with an initial scale height of 75\,pc for a total age of 1\,Gyr or less, resulting in an age-average scale height of 230\,pc for the local sample. The simulation also assumes a limiting {\it Gaia} magnitude of $G = 20$. We have repeated the simulation by artificially multiplying and dividing the disc scale height by a factor of two, respectively. The results are presented in Fig.\,\ref{completeness3}, where the observed {\it Gaia} space density agrees remarkably well with our standard model and an age-average disc scale height of 230\,pc. The simulation naturally explains the smaller space density within 100\,pc compared to 20\,pc without the need to invoke the completeness of {\it Gaia}, for which the selection function is not expected to change remarkably between these distances. 

\section{Conclusion}
We retrieved the available \textit{Gaia} DR2 data for $\simeq 24\,000$ spectroscopically confirmed white dwarfs from SDSS, and analysed the properties and distribution of these objects in the Hertzsprung-Russell diagram to define a reliable method to select high-confidence white dwarf candidates from \textit{Gaia} DR2. After defining several quality cuts to remove objects with poor \textit{Gaia} measurements, we find that no simple selection relying solely on \textit{Gaia} colour and absolute magnitude can separate white dwarfs from contaminant objects without excluding a significant number of known white dwarfs. We therefore make use of the distribution in $\bp-\rp$ colour and \Gabs\ of a sample of spectroscopically confirmed white dwarfs and contaminants from SDSS to calculate \textit{probabilities of being a white dwarf} (\Pwd) for all \textit{Gaia} objects in our sample. This results in a total of \finsize\ objects with calculated \Pwd\ from which it is possible to select a sample of $\simeq 260\,000$ high-confidence white dwarf candidates. The \Pwd\ values, coupled with \textit{Gaia} quality flags, can be used to flexibly select samples of white dwarfs with varying degree of completeness and contamination according to one's specific goals. For general purpose we recommend a cut at $\Pwd>0.75$, which we estimate includes 95~per cent of all the white dwarfs in the total sample, with minimal level of contamination ($\simeq 4$ per cent).  
We also provide stellar parameters (\Teff, $\log g$ and mass) for a subsample of 225\,370 candidates that have \textit{Gaia} parallax and photometric measurements precise enough to achieve a reliable fit to our adopted models. We find the atmospheric parameters obtained fitting only \textit{Gaia} observations to be in good agreement with those obtained using SDSS and Pan-STARRS photometry.

We further characterised the \textit{Gaia} sample of white dwarfs by visually inspecting the observed cooling sequence in the H-R diagram of representative samples of spectroscopically confirmed stellar remnants from the SDSS. We identify a number of sub-structures in the white dwarf cooling tracks, some of which are the result of different spectral types and others that remain unexplained.

We have used a newly constructed sample of SDSS white dwarf candidates selected on the basis of their colours and proper motions to estimate the overall completeness of our \textit{Gaia} catalogue of white dwarf candidates. We found the catalogue to be between 60 and 85 per cent complete for white dwarfs with $G$ $\leq20$ and \Teff\ $> 7000$\,K, at high Galactic latitudes ($|b|>20^{\circ}$). 

The presented \textit{Gaia} catalogue represents the first step towards a homogeneous all-sky census of all white dwarfs, and to fully explore the rich scientific potential of this sample, spectroscopic follow-up will ultimately be needed to study these objects in detail.
The $P_{\mathrm{WD}}$ values that we derived allow to tailor future spectroscopic campaigns prioritising efficiency for single target observations or completeness in large scale surveys.
With large multi-fibre spectroscopic facilities approaching first light in both hemispheres (e.g., WEAVE, 4MOST, DESI, SDSS-V, \citealt{weave14-1,4most14-1,DESI16-1,sdssv17-1}), our catalogue represents a key resource for future white dwarf studies.

\section*{Acknowledgements}
The research leading to these results has received funding from the European Research Council under the European Union's Horizon 2020 research and innovation programme n. 677706 (WD3D) and the European Union's Seventh Framework Programme (FP/2007- 2013) / ERC Grant Agreements n. 320964 (WDTracer). Additional funding was provided by STFC via grant ST/P000495/1. R.~R. acknowledges funding by the German Science foundation (DFG) through grants HE1356/71-1 and IR190/1-1.

This work has made use of data from the European Space Agency (ESA) mission {\it Gaia} (\url{https://www.cosmos.esa.int/gaia}), processed by the {\it Gaia} Data Processing and Analysis Consortium (DPAC, \url{https://www.cosmos.esa.int/web/gaia/dpac/consortium}). Funding for the DPAC has been provided by national institutions, in particular the institutions participating in the {\it Gaia} Multilateral Agreement. 

The work presented in this article made large use of TOPCAT and STILTS Table/VOTable Processing Software \citep{topcat-1}.

This work has made use of observations from the SDSS-III, funding for which has been provided by the Alfred P. Sloan Foundation, the Participating Institutions, the National Science Foundation, and the U.S. Department of Energy Office of Science. The SDSS-III web site is http://www.sdss3.org/.

We thank Alberto Rebassa Mansergas for his constructive feedback.
\bibliographystyle{mnras}
\bibliography{aamnem99,aabib1,aabib2,aabib_new,aabib_tremblay}

\appendix
\section{The deep SDSS-GPS1 comparison sample}
\label{app_A}
\begin{figure}
\includegraphics[width=\columnwidth]{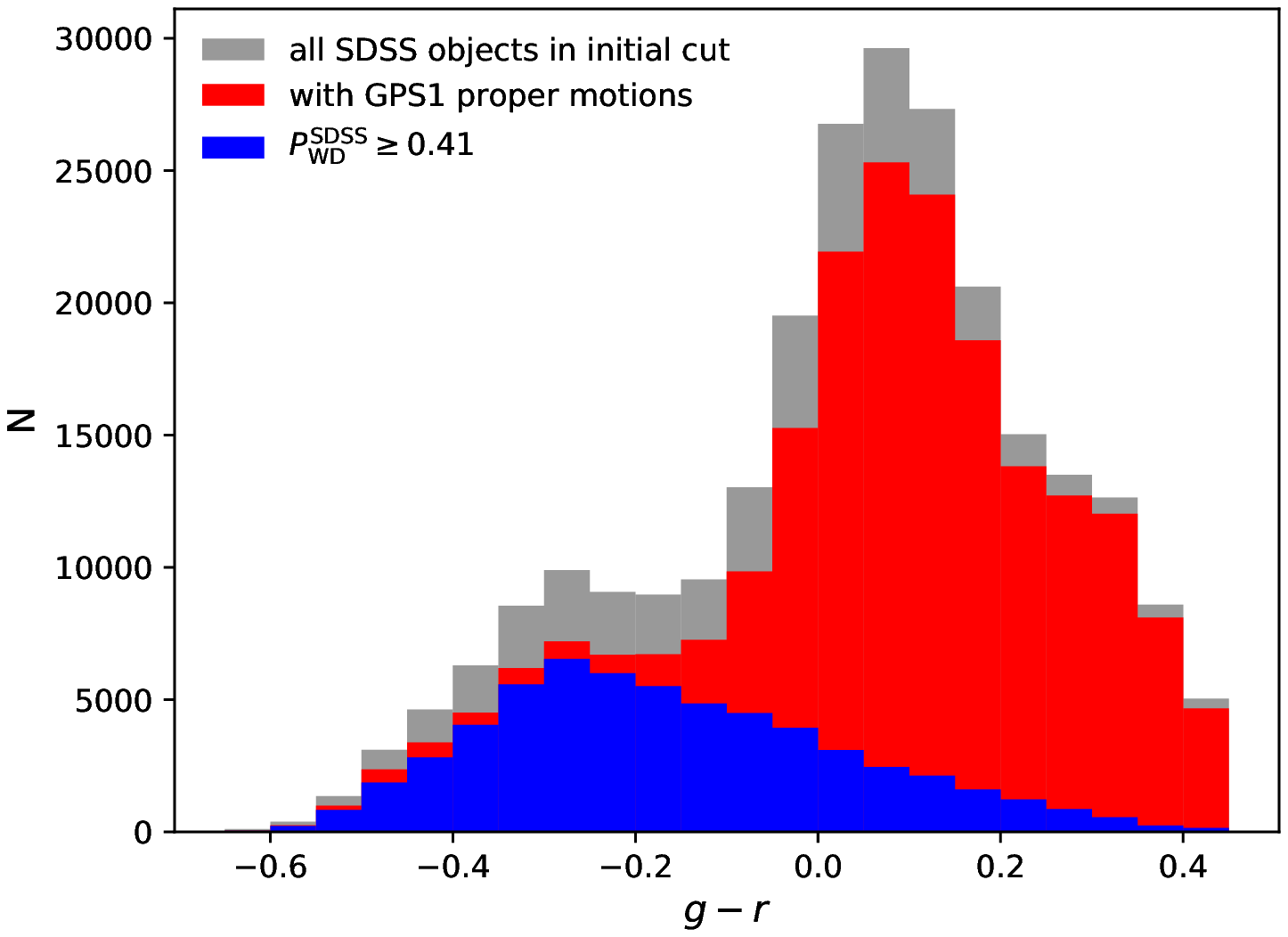}
\caption{\label{hist_compare} Distribution in $g-r$ of the SDSS sources selected for the comparison sample. The number of objects for which no GPS1 proper motion could be retrieved is not uniform across the colour axis.}
\end{figure}

Our original catalogue of SDSS white dwarf candidates published in \citet{gentilefusilloetal15-1} relied on SDSS DR10 proper motions and was limited to objects brighter than $g=19$. In order to create a  comparison sample matching the depth of our \textit{Gaia} catalogue we needed to extend the sample of SDSS white dwarfs candidates to fainter magnitudes. As SDSS proper motions quickly become unreliable past $g=19$ we adopted proper motions from the Gaia-PS1-SDSS (GPS1) Catalog \citep{tianetal17-1}.
Following the same $ugriz$ colour selection as described in \citet{gentilefusilloetal15-1} and \citet{gentilefusilloetal17-1} we selected 263\,944 blue SDSS point sources with $g\leq20.1$. This colour cut limits the sample to only white dwarfs with \Teff\ $>7000$\,K. Large areas of the sky at RA $<12$ are entirely missing in the GPS1 catalogues and no proper motions could be retrieved for objects at these locations. In order to circumvent the effects of these gaps in GPS1 we further limit our comparison sample to SDSS sources with RA $>12$ before carrying out the cross match with GPS1. This brings the number of objects in the sample to 253\,640. We cross matched the positions of these objects with GPS1 to retrieve their proper motions. Coordinates in GPS1 are provided in epoch J2010 while SDSS observations were collected between 2000 and 2008. Since high proper motions objects like white dwarfs can move significantly over these time scales, we carried out our cross-match accounting for this epoch difference following the method described in \citet{gentilefusilloetal17-1} and finally retrieved proper motions for 211\,988 of the 253\,640 SDSS objects. Using SDSS colours and GPS1 proper motions we then calculated a reduced-proper-motion based \emph{probability of being a white dwarf} $P_\mathrm{WD}^\mathrm{SDSS}$ for all our objects using the method described in \citet{gentilefusilloetal15-1}. These $P_\mathrm{WD}^\mathrm{SDSS}$ values are different and unrelated 
to the \textit{Gaia}-based \Pwd\ values presented in Section\,\ref{selection}.
Using our training set of spectroscopically confirmed SDSS white dwarfs and contaminants, we can calculate completeness (ratio of the number of selected white dwarfs to the total number of white dwarfs) and an efficiency (ratio of the number of selected white dwarfs to the total number of objects selected)  for different threshold values of $P_\mathrm{WD}^\mathrm{SDSS}$. For example, selecting all objects with  $P_\mathrm{WD}^\mathrm{SDSS}\geq 0.41$ results in a sample 97~per cent complete with an efficiency of 93~per cent. This also allows us to estimate that the entire sample of objects for which we calculated  $P_\mathrm{WD}^\mathrm{SDSS}$ contains $\simeq 56\,600$ white dwarfs.
However, we could retrieve GPS1 proper motions (and so calculate $P_\mathrm{WD}^\mathrm{SDSS}$ values) only for $\simeq84$\,per cent  of the SDSS objects within our initial colour and RA cut. Additionally, there appears to be a colour ($g-r$) dependence in the number of objects for which no proper motion was found (Fig.\,\ref{hist_compare}). When combining this effect with the distribution of white dwarfs in $g-r$ and the efficiency of our $P_\mathrm{WD}^\mathrm{SDSS}$ cut in different bins of colour-space, we find that on average up to 25~per cent of white dwarfs may not have been included in the sample as a result of not having a proper motion in the GPS1 catalogue (Fig.\,\ref{hist_compare}). We therefore conclude that our deep SDSS comparison sample of objects with calculated $P_\mathrm{WD}^\mathrm{SDSS}$ only includes 75\,per cent of all the white dwarfs in the SDSS footprint with RA $>$ 12, $g\leq20.1$ and $\Teff>7000$\,K. Nonetheless, we can estimate that an additional $\simeq 14\,000$ white dwarfs are among the objects initially included in our SDSS colour-cut, but which have no proper motion in GPS1.

\end{document}